\documentclass{article}
\usepackage[utf8]{inputenc}
\usepackage{authblk}
\usepackage{xspace}
\usepackage{url}
\usepackage[table, svgnames, dvipsnames]{xcolor}
\usepackage{mathptmx}
\usepackage{microtype}
\usepackage{booktabs}
\usepackage{siunitx}
\usepackage{enumitem}
\usepackage{multicol, caption}
\usepackage{nonfloat}
\usepackage{makecell}
\usepackage{amsmath}
\usepackage{graphicx}
\usepackage{lineno}
\usepackage{pifont}
\usepackage[labelformat=simple]{subcaption}
\usepackage[justification=justified]{caption}

\newcommand{\cmark}{\ding{51}}%
\newcommand{\xmark}{\ding{55}}%

\usepackage[margin={2cm, 2.1cm}]{geometry}

\usepackage[style=nature,doi=true,url=true]{biblatex}
\title{Privacy-preserving Artificial Intelligence Techniques in Biomedicine}
\author[1]{Reihaneh Torkzadehmahani}
\author[1]{Reza Nasirigerdeh}
\author[1]{David B. Blumenthal}
\author[1, 2]{Tim Kacprowski}
\author[1]{Markus List}
\author[1]{Julian Matschinske}
\author[1]{Julian Späth}
\author[1]{Nina Kerstin Wenke}

\author[3]{Béla Bihari}
\author[4]{Tobias Frisch}
\author[4]{Anne Hartebrodt}
\author[5]{Anne-Christin Hauschild}
\author[5]{Dominik Heider}
\author[6]{Andreas Holzinger}
\author[7]{Walter Hötzendorfer}
\author[7]{Markus Kastelitz}
\author[8]{Rudolf Mayer}
\author[9]{Cristian Nogales}
\author[8]{Anastasia Pustozerova}
\author[4]{Richard Röttger}
\author[9]{Harald H.H.W. Schmidt}
\author[10]{Ameli Schwalber}
\author[7]{Christof Tschohl}
\author[10]{Andrea Wohner}

\author[1, 4]{Jan Baumbach}

\affil[1]{Chair of Experimental Bioinformatics, Technical University of Munich, Freising, Germany}
\affil[2]{Division of Data Science in Biomedicine, Peter L. Reichertz Institute for Medical Informatics of TU Braunschweig and Hannover Medical School, Braunschweig, Germany}
\affil[3]{Gnome Design SRL, Sfântu Gheorghe, Romania}
\affil[4]{Institute of Mathematics and Computer Science, University of Southern Denmark, Odense, Denmark }
\affil[5]{Department of Mathematics and Computer Science, Philipps-University of Marburg, Marburg, Germany}
\affil[6]{Institute for Medical Informatics/Statistics, Medical University Graz, Graz, Austria}
\affil[7]{Research Institute AG \& Co KG, Vienna, Austria}
\affil[8]{SBA Research Gemeinnützige GmbH, Vienna, Austria}
\affil[9]{Department of Pharmacology and Personalised Medicine, MeHNS, FHML, Maastricht University, Maastricht, the Netherlands }
\affil[10]{Concentris Research Management GmbH, Fürstenfeldbruck, Germany }

\date{}

\begin{document}
    \maketitle
    \section*{Abstract}
      Artificial intelligence (AI) has been successfully applied in numerous scientific domains. In biomedicine, AI has already shown tremendous potential, e.g. in the interpretation of next-generation sequencing data and in the design of clinical decision support systems. However, training an AI model on sensitive data raises concerns about the privacy of individual participants. For example, summary statistics of a genome-wide association study can be used to determine the presence or absence of an individual in a given dataset. This considerable privacy risk has led to restrictions in accessing genomic and other biomedical data, which is detrimental for collaborative research and impedes scientific progress. Hence, there has been a substantial effort to develop AI methods that can learn from sensitive data while protecting individuals' privacy. This paper provides a structured overview of recent advances in privacy-preserving AI techniques in biomedicine. It places the most important state-of-the-art approaches within a unified taxonomy and discusses their strengths, limitations, and open problems. As the most promising direction, we suggest combining federated machine learning as a more scalable approach with other additional privacy preserving techniques. This would allow to merge the advantages to provide privacy guarantees in a distributed way for biomedical applications. Nonetheless, more research is necessary as hybrid approaches pose new challenges such as additional network or computation overhead.

    \begin{multicols}{2}
    \section*{Introduction}
         Artificial intelligence (AI) strives to emulate human mind and to solve complex tasks by learning from available data. For many complex tasks, AI already surpasses humans in terms of accuracy, speed and cost.  Recently, the rapid adoption of AI and its subfields, specifically machine learning and deep learning, has led to substantial progress in applications such as autonomous driving \cite{Schwarting_2018}, text translation \cite{gehring2017convolutional} and voice assistance \cite{xiong2018microsoft}.  At the same time, AI is becoming essential in biomedicine, where big data in healthcare necessitates techniques that help scientists to gain understanding from it \cite{HolzingerKieseWeipplTjoa:2018:trends}.
        
        Success stories such as acquiring the compressed representation of drug-like molecules \cite{gomez2018automatic}, modeling the hierarchical structure and function of a cell \cite{ma2018using} and translating magnetic resonance images to computed tomography \cite{nie2018medical} using deep learning models illustrate the remarkable performance of these AI approaches. AI has not only achieved remarkable success in analyzing genomic and biomedical data \cite{hosny2018artificial, beam2018big, yu2018artificial, michael2018visible, chen2018rise, wainberg2018deep, min2017deep, litjens2017survey, shen2017deep, jiang2017artificial,libbrecht2015machine}, but has also surpassed humans in applications such as sepsis prediction \cite{nemati2018interpretable}, malignancy detection on mammography \cite{teare2017malignancy} and mitosis detection in breast cancer \cite{veta2015assessment}.
        
        Despite these AI-fueled advancements, important privacy concerns have been raised regarding the individuals who contribute to the datasets. While taking care of the confidentiality of sensitive biological data is crucial \cite{naveed2015privacy}, several studies showed that AI techniques often do not maintain data privacy \cite{shokri2017membership, papernot2018sok, zhang2016understanding, zhang2020secret}. For example, attacks known as membership inference can be used to infer an individual's membership by querying over the dataset \cite{shringarpure2015privacy} or the trained model \cite{shokri2017membership}, or by having access to certain statistics about the dataset \cite{homer2008resolving, harmanci2018analysis, wang2009learning}. Homer et al. \cite{homer2008resolving} showed that under some assumptions, an adversary (an attacker who attempts to invade data privacy) can use the statistics published as the results of genome-wide association studies (GWAS) to find out if an individual was a part of the study.  Another example of this kind of attack was demonstrated by attacks on Genomics Beacons \cite{beacon, shringarpure2015privacy}, in which an adversary could determine the presence of an individual in the dataset by simply querying the presence of a particular allele. Moreover, the attacker could identify the relatives of those individuals and obtain sensitive disease information \cite{wang2009learning}.  Besides targeting the training dataset, an adversary may attack a fully-trained AI model to extract individual-level membership by training an adversarial inference model that learns the behaviour of the target model \cite{shokri2017membership}.
        
        As a result of the aforementioned studies, health research centers such as the National Institutes of Health (NIH) as well as hospitals have restricted access to the pseudonymized data \cite{zerhouni2008protecting, erlich2014routes, naveed2015privacy}. Furthermore, data privacy laws such as those enforced by the Health Insurance Portability and Accountability Act (HIPAA), and the Family Educational Rights and Privacy Act (FERPA) in the US as well as the EU General Data Protection Regulation (GDPR) restrict the use of sensitive data \cite{GDPR,cohen2020towards}. 
        Consequently, getting access to these datasets requires a lengthy approval process, which significantly impedes collaborative research. Therefore, both industry and academia urgently need to apply privacy-preserving techniques to respect individual privacy and comply with these laws.
        
        This paper provides a systematic overview over various recently proposed privacy-preserving AI techniques in biomedicine, which facilitate the collaboration between health research institutes. Several efforts exist to tackle the privacy concerns in several domains, some of which have been examined in a couple of surveys~\cite{aziz2019privacy, xu2019federated, Kaissis_2020}. Aziz \textit{et al.}~\cite{aziz2019privacy} investigated previous studies which employed differential privacy and cryptographic techniques for human genomic data.  Kaissis \textit{et al.}\cite{Kaissis_2020} briefly reviewed federated learning, differential privacy and cryptographic techniques applied in medical imaging. Xu \textit{et al.}~\cite{xu2019federated} surveyed general solutions to challenges in federated learning including communication efficiency, optimization, as well as privacy and discussed  possible applications including a few examples in healthcare. Compared to \cite{aziz2019privacy} and \cite{Kaissis_2020}, this paper covers a broader set of privacy preserving techniques including federated learning and hybrid approaches. In contrast to \cite{xu2019federated} we additionally discuss cryptographic techniques and differential privacy approaches and their applications in biomedicine. Moreover, this survey covers a wider range of studies that employed different privacy-preserving techniques in genomics and biomedicine and compares the approaches using different criteria such as privacy, accuracy and efficiency.
         \noindent   
            \begin{figure*}[!htb]
            \begin{minipage}{0.5\textwidth}
                 \centering
                 \includegraphics[width=1\textwidth]{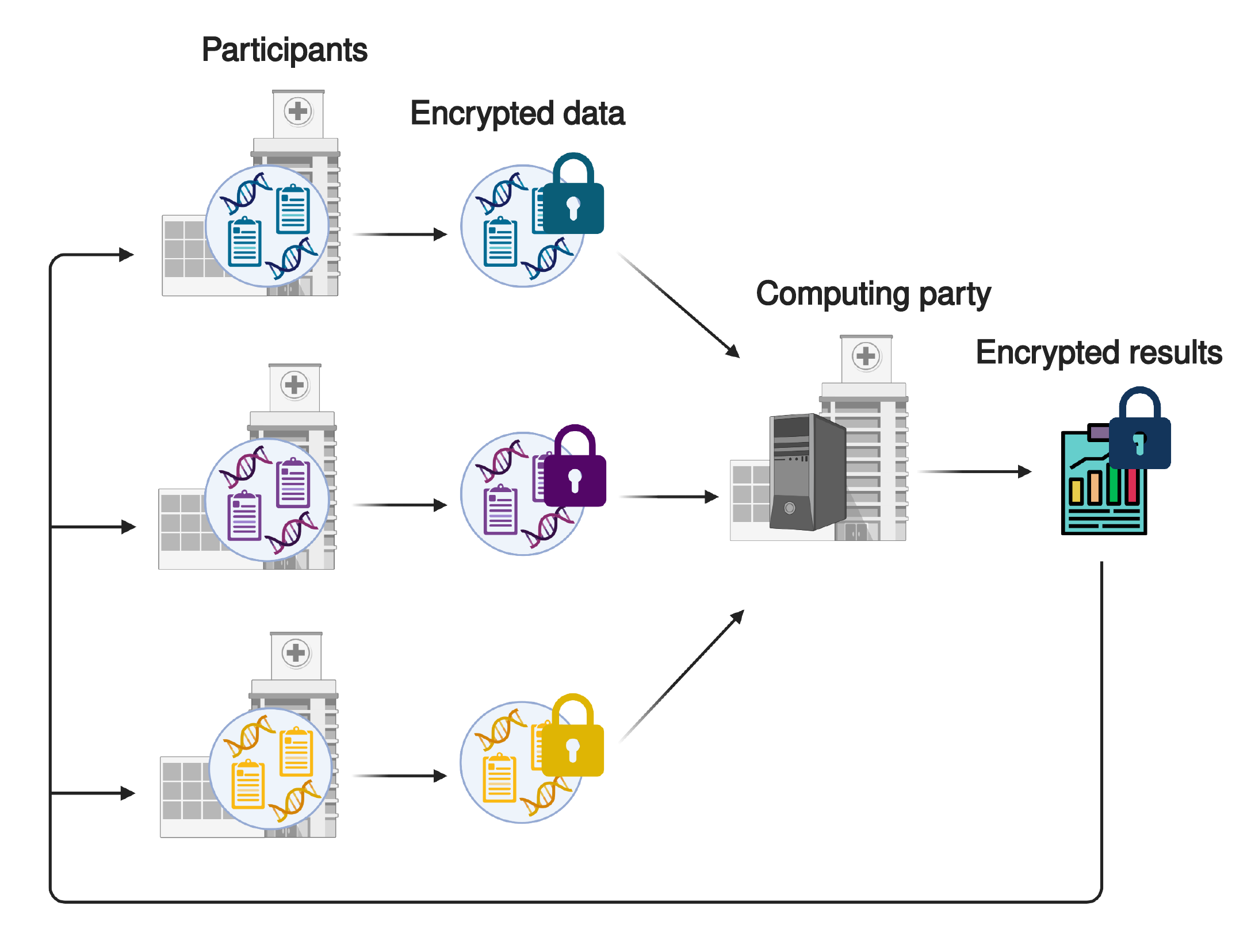}
                 \subcaption[]{) Homomorphic encryption}
                 \medskip
            \end{minipage}
            \begin{minipage}{0.5\textwidth}
                 \centering
                 \includegraphics[width=1\textwidth]{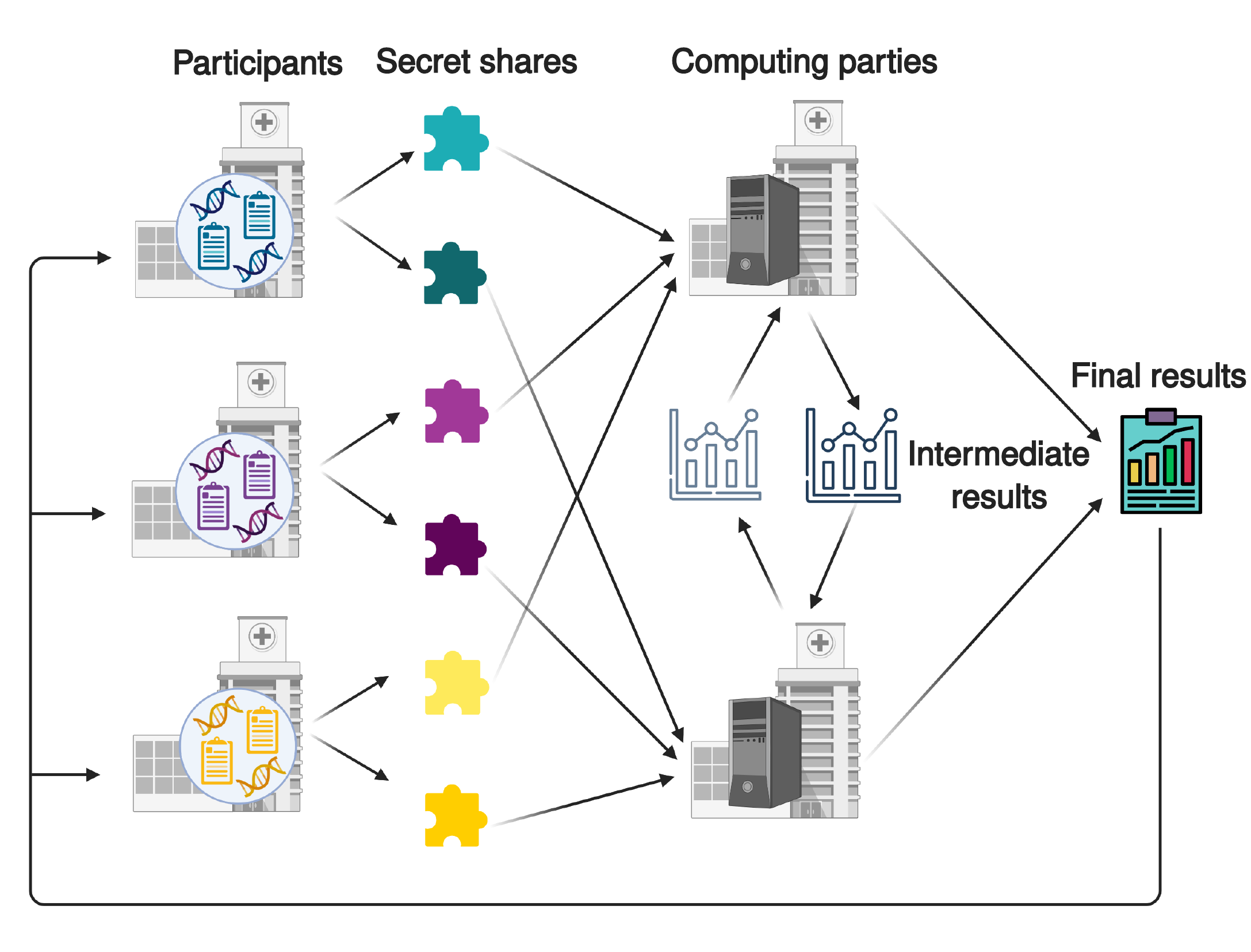}
                 \subcaption[]{) Secure multiparty computation}
                 \medskip
            \end{minipage}
            \begin{minipage}{0.5\textwidth}
                 \centering
                 \includegraphics[width=1\textwidth]{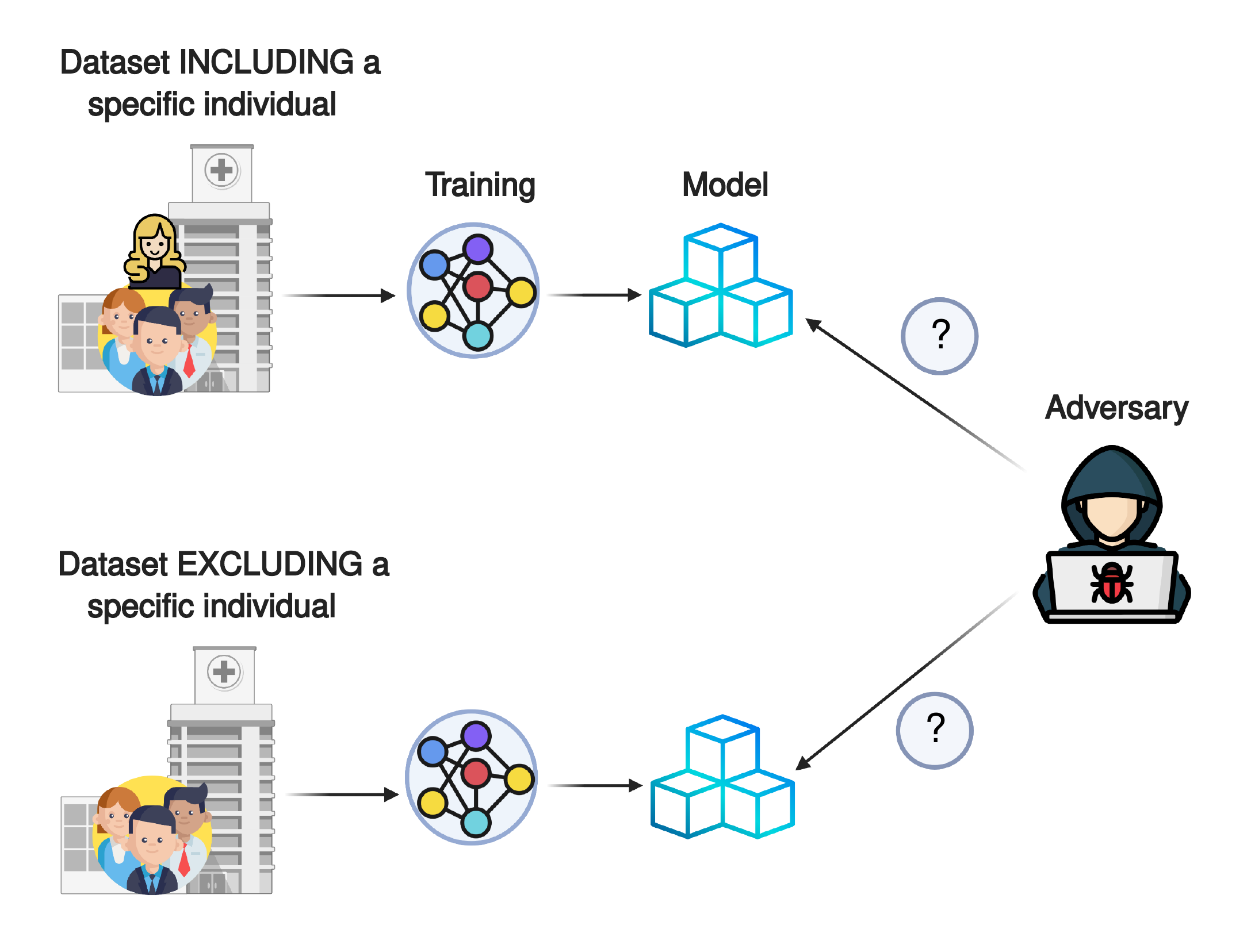}
                 \subcaption[]{) Differential privacy}
            \end{minipage}
            \begin{minipage}{0.5\textwidth}
                 \centering
                 \includegraphics[width=1\textwidth]{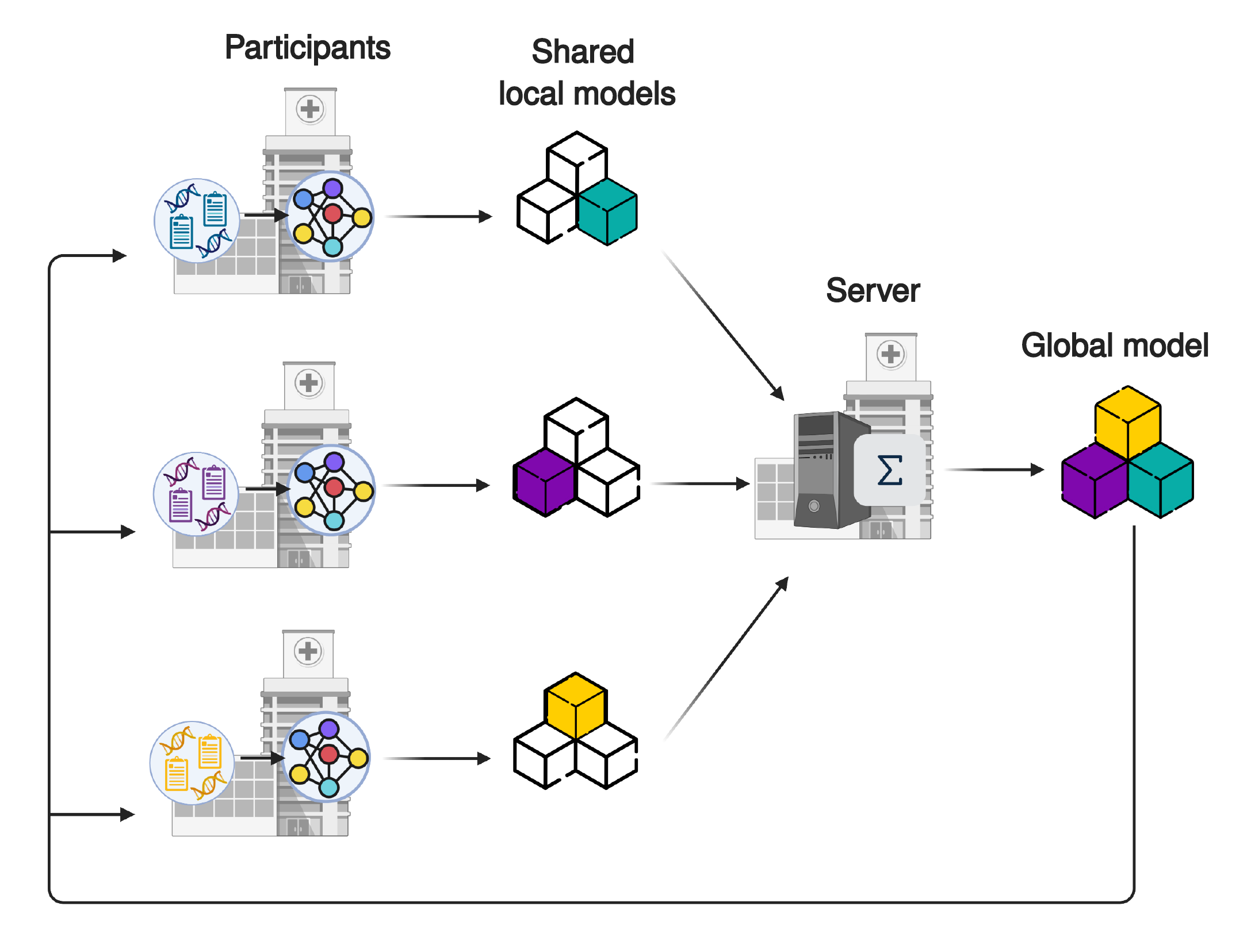}
                 \subcaption[]{) Federated learning }
            \end{minipage}
             \caption{Different privacy-preserving AI techniques: (a) \textbf{homomorphic encryption}, where the participants encrypt the private data and share it with a computing party, which computes the aggregated results over the encrypted data from the participants;
             (b) \textbf{secure multiparty computation} in which each participant shares a separate, different secret with each computing party; the computing parties calculate the intermediate results, secretly share them with each other, and  aggregate all intermediate results to obtain the final results;
             (c) \textbf{differential privacy}, which ensures the models trained on datasets including and excluding a specific individual look statistically indistinguishable to the adversary; (d) \textbf{federated learning}, where each participant downloads the global model from the server, computes the local model given its private data and the global model, and finally sends its local model to the server for aggregation and for updating the global model.}
            \label{fig:techniques}
            \end{figure*}

        The presented approaches are divided into four categories: cryptographic techniques, differential privacy, federated learning, and hybrid approaches. First, we describe how cryptographic techniques\,---\,in particular, homomorphic encryption (HE) and secure multiparty computation (SMPC)\,---\,ensure secrecy of sensitive data by carrying out computations on encrypted/anonymous biological data. Next, we illustrate the differential privacy approach and its capability in quantifying individuals’ privacy in published summary statistics of, for instance, GWAS data and deep learning models trained on clinical data. Then, we elaborate on federated learning, which allows health institutes to train AIs locally and to share only selected parameters without sensitive data with a coordinator, who aggregates them and builds a global model. Following that, we discuss hybrid approaches which enhance data privacy by combining federated learning with other privacy-preserving techniques. We elaborate on the strengths and drawbacks of each approach as well as its applications in biomedicine. More importantly, we provide a comparison among the approaches with respect to different criteria such as computational and communication efficiency, accuracy, and privacy. Finally, we discuss the most realistic approaches from a practical viewpoint and provide a list of open problems and challenges that remain for the adoption of these techniques in real-world biomedical applications.

        Our review of privacy-preserving AI techniques in biomedicine yields the following main insights: First, cryptographic techniques such as HE and SMPC, which follow the paradigm of ``bring data to computation``, are not  computationally efficient and do not scale well to large biomedical datasets. Second, federated learning follows the paradigm  of ``bring computation to data`` is a more scalable approach. However, its network communication efficiency is still an open problem and it does not provide privacy guarantees. Third, hybrid approaches that combine cryptographic techniques or differential privacy with federated learning are the most promising privacy-preserving AI techniques for biomedical applications, because they promise to combine the scalability of federated learning with the privacy guarantees of cryptographic techniques or differential privacy.
        
        \section*{Cryptographic Techniques}\label{sec:crypto_techniques}
        
            In biomedicine and GWAS in particular, cryptographic techniques have been used to collaboratively compute result statistics while preserving data privacy \cite{cho2018secure, bonte2018towards, jagadeesh2019keeping, kim2015private, lauter2014private, lu2015privacy, zhang2015foresee, kamm2013new, constable2015privacy, zhang2015secure, mohassel2017secureml, chen2017princess, hasan2018secure, sadat2018safety}. These cryptographic approaches are based on HE \cite{gentry2009fully}  or SMPC \cite{cramer2015secure}. HE-based approaches share three steps (Figure~\ref{fig:techniques}a): 
             \begin{enumerate}
                \item Participants (e.g. hospitals or medical centers) encrypt their private data and send the encrypted data to a computing party.
                \item  The computing party calculates the statistics over the encrypted data and shares the statistics (which are encrypted) with the participants.
                \item The participants access the results by decrypting them.
            \end{enumerate}

            In SMPC,  there are multiple participants as well as a couple of computing parties which perform computations on secret shares from the participants. 
            Given $M$ participants and $N$ computing parties, SMPC-based approaches follow three steps (Figure~\ref{fig:techniques}b):
            \begin{enumerate}
                \item Each participant sends a separate and different secret to each of the $N$ computing parties.
                
                \item Each computing party computes the intermediate results on the $M$ secret shares from the participants and shares the intermediate results with the other $N-1$ computing parties.
                \item Each computing party aggregates the intermediate results from all computing parties including itself to calculate the final (global) results. In the end, the final results computed by all computing parties are the same and can be shared by the participants.
            \end{enumerate}
            
            To clarify the concepts of secret sharing~\cite{shamir1979secretshare} and multi-party computation, consider a scenario with two participants $P_1$ and $P_2$ and two computing parties $C_1$ and $C_2$~\cite{jagadeesh2019smpcscenario}. $P_1$ and $P_2$ possess the private data $X$ and $Y$, respectively. The aim is to compute $X+Y$, where neither $P_1$ nor $P_2$ reveals its data to the computing parties. To this end, $P_1$ and $P_2$ generate random numbers $R_X$ and $R_Y$, respectively; $P_1$ reveals $R_X$ to $C_1$ and $(X - R_X)$ to $C_2$; likewise, $P2$ shares $R_Y$ with $C_1$ and $(Y - R_Y)$ with $C_2$; $R_X$, $R_Y$, $(X - R_X)$ and $(Y - R_Y)$ are secret shares. $C_1$ computes $(R_X + R_Y)$ and sends it to $C_2$ and $C_2$ calculates $(X - R_X) + (Y - R_Y)$ and reveals it to $C_1$. Both $C_1$ and $C_2$ add the result they computed to the result each obtained from the other computing party. The sum is in fact $(X + Y)$, which can be shared with $P_1$ and $P_2$.
                     
            Notice that to preserve data privacy, the computing parties $C_1$ and $C_2$ must be non-colluding. That is, $C_1$ must not send $R_X$ and $R_Y$ to $C_2$ and $C_2$ must not share $(X - R_X)$ and $(Y - R_Y)$ with $C_1$. Otherwise, the computing parties can compute $X$ and $Y$, revealing the participants' data. In general, in a SMPC with $N$ computing parties, data privacy is protected as long as at most $N-1$ computing parties collude with each other. The larger $N$, the stronger the privacy but the higher the communication overhead and processing time.
            
            Most studies use HE or SMPC to develop secure, privacy-aware algorithms applicable to GWAS data. Kim \textit{et al.} \cite{kim2015private} and Lu \textit{et al.} \cite{lu2015privacy} implemented a secure $\chi^2$ test and Lauter \textit{et al.} \cite{lauter2014private} developed privacy-preserving versions of common statistical tests in GWAS, such as the Pearson goodness of fit test, tests for linkage disequilibrium, and the Cochran Armitage trend test using HE. Kim \textit{et al.} \cite{kim2018secure} and Morshed \textit{et al.} \cite{morshed2018parallel} presented a HE-based secure logistic and linear regression algorithms for GWAS data, respectively. Zhang \textit{et al.}~\cite{zhang2015secure}, Constable \textit{et al.}~\cite{constable2015privacy}, and Kamm \textit{et al.}~\cite{kamm2013new} developed a SMPC-based secure $\chi^2$ test. Shi \textit{et al.}~\cite{shi2016smpclogistic} implemented a privacy-preserving logistic regression and Bloom~\cite{bloom2019smpclinear} proposed a secure linear regression based on SMPC for GWAS data. Cho \textit{et al.}~\cite{cho2018secure} introduced a SMPC-based framework to facilitate quality control and population stratification correction for large-scale GWAS and argued that their framework is scalable to one million individuals and half million single nucleotide polymorphisms (SNPs). 
            
            There are also other types of encryption techniques such as somewhat homomorphic encryption (SWHE)~\cite{gentry2009fully}, which are  employed to address privacy issues in genomic applications such as outsourcing genomic data computation to the cloud, and are not the main focus of this review. For more details, we refer to the comprehensive review by Mittos \textit{et al.} \cite{mittos2019systematizing}.
            
            Despite the promises of HE/SMPC-based privacy-preserving algorithms (Table~\ref{tab:Crypto_dp}), the road for the wide adoption of HE/SMPC-based algorithms in genomics and biomedicine is long \cite{berger2019cryptlimit}. The major limitations of HE are few supported operations and computational overhead ~\cite{chialva2018helimit}. HE supports only addition and multiplication operations, and as a result, developing complex AI models with non-linear operations such as deep neural networks (DNNs) using HE is very challenging. Moreover, HE incurs remarkable computational overhead since it performs operations on encrypted data. Although SMPC is more efficient than HE from a computational perspective, it still suffers from high computational overhead~\cite{alexandru2020smpclimits}, which comes from processing secret shares from a large number of participants or large amount of data by a few number of computing parties. 
            
         \begin{table*}[ht]
                \centering
                \caption{Literature for \textbf{cryptographic techniques} and \textbf{differential privacy} in genomics and biomedicine. HE: homomorphic encryption, SMPC: secure multiparty computation, DP: differential privacy}
                \medskip
                \begin{tabular}{lSlll}\toprule
                    \textbf{Authors}   & \textbf{Year} & \textbf{Privacy Technique}  & \textbf{Model}  & \textbf{Application}  \\ 
                    \midrule
                    \rowcolor{Gainsboro!90}
                    \text{Kim \textit{et al.} \cite{kim2015private}}  & \text{2015}  & \text{HE}  & \text{\makecell[l]{$\chi^2$ statistics \\ minor allele frequency \\ Hamming Distance \\
                    Edit distance}}  & \text{\makecell[l]{genetic associations\\ DNA comparison} }  \\
                    
                    \text{ Lu \textit{et al.} \cite{lu2015privacy}} & \text{2015} & \text{HE}  & \text{\makecell[l]{$\chi^2$ statistics \\ $D^\prime$ measure}}  & \text{\makecell[l]{genetic associations} }  \\

                    \rowcolor{Gainsboro!60}\text{Lauter \textit{et al.} \cite{lauter2014private}}  & \text{2014} & \text{HE}  &  \text{\makecell[l]{$D^\prime$ and $r^2$ measures \\ Pearson goodness-of-fit \\ expectation maximization \\ Cochran-Armitage}}  & \text{genetic associations}  \\
                    
                    \text{Kim \textit{et al.} \cite{kim2018secure} }   & \text{2018} & \text{HE}  & \text{logistic regression }  & \text{medical decision making}  \\
                    
                    \rowcolor{Gainsboro!60}
                    \text{Morshed \textit{et al.} \cite{morshed2018parallel}}  & \text{2018} & \text{HE}  &  \text{linear regression}  & \text{medical decision making} \\
                    
                    \midrule
                    \text{Kamm \textit{et al.}~\cite{kamm2013new}}  & \text{2013} & \text{SMPC}  & \text{$\chi^2$ statistics}  & \text{genetic associations}  \\
                    
                     \rowcolor{Gainsboro!60}\text{\makecell[l]{Constable \textit{et al.}~\cite{constable2015privacy} \\ Zhang \textit{et al.}~\cite{zhang2015secure}}}   & \text{\makecell[l]{2015 \\2015 }} & \text{SMPC}  & \text{\makecell[l]{$\chi^2$ statistics \\ minor allele frequency}}  & \text{genetic associations}  \\
                     
                    \text{Shi \textit{et al.}~\cite{shi2016smpclogistic}}   & \text{2016} & \text{SMPC}  &  \text{logistic regression}  &  \text{genetic associations}  \\ 
                    
                    \rowcolor{Gainsboro!60}\text{Bloom~\cite{bloom2019smpclinear}}   & \text{2019} & \text{SMPC}  &  \text{linear regression} &  \text{genetic associations}   \\
                    
                    \text{Cho \textit{et al.}~\cite{cho2018secure}}   & \text{2018} & \text{SMPC}  & \text{\makecell[l]{quality control \\ population stratification}} & \text{genetic associations} \\
                    \midrule
                     \rowcolor{Gainsboro!60}\text{Johnson \textit{et al.}~\cite{johnson2013privacy}} & \text{2013} &  \text{DP}  & \text{\makecell[l]{distance-score mechanism \\ p-value and $\chi^2$ statistics}} & \text{querying genomics databases} \\
                     
                     \text{Cho \textit{et al.}~\cite{cho2020privacy}} & \text{2020} &  \text{DP}  & \text{\makecell[l]{truncated $\alpha$ -geometric mechanism}} & \text{querying biomedical databases} \\
                     
                    \rowcolor{Gainsboro!60}\text{Aziz \textit{et al.}~\cite{al2017aftermath}} & \text{2017} &  \text{DP}  & \text{\makecell[l]{eliminating random positions \\ biased random response}} & \text{querying genomics databases} \\
                    
                    \text{\makecell[l]{Han \textit{et al.}~\cite{han2019differential} \\ Yu \textit{et al.}~\cite{yu2014differentially}}}  & \text{\makecell[l]{2019 \\2014}} &  \text{DP}  & \text{logistic regression} & \text{genetic associations}\\
                    
                    \rowcolor{Gainsboro!60}\text{Honkela \textit{et al.}~\cite{honkela2018efficient}} &  \text{2018} &  \text{DP}  &\text{bayesian linear regression} & \text{drug sensitivity prediction}\\
                    
                    \text{Simmons \textit{et al.}~\cite{simmons2016enabling}}  & \text{2016} &  \text{DP}  & \text{\makecell[l]{ EIGENSTRAT \\ linear mixed model}} & \text{genetic associations}\\
                
                    \rowcolor{Gainsboro!60}\text{Simmons \textit{et al.}~\cite{simmons2016realizing}}  & \text{2016} &  \text{DP}  &\text{nearest neighbor optimization} & \text{\text{genetic associations}}\\
                    
                    \text{\makecell[l]{Fienberg \textit{et al.}~\cite{fienberg2011privacy} \\ Uhlerop \textit{et al.}~\cite{uhlerop2013privacy} \\ Yu \textit{et al.}~\cite{yu2014scalable} \\ Wang \textit{et al.}~\cite{wang2014differentially}}}  & \text{\makecell[l]{2011 \\ 2013 \\ 2014 \\ 2014}} &  \text{DP}  & \text{\makecell[l]{statistics such as p-value, \\$\chi^2$ and contingency table }}  & \text{genetic associations}  \\
                    
                   \rowcolor{Gainsboro!60}\text{Abay \textit{et al.}~\cite{abay2018privacy}}  & \text{2018} &  \text{DP}  &\text{deep autoencoder} & \text{\text{generating artificial biomedical data}}\\
                   
                   \text{Beaulieu \textit{et al.}~\cite{beaulieu2019privacy}}  & \text{2019} &  \text{DP}  & \text{GAN} & \text{\text{simulating SPRINT trial}}\\
                   
                   \rowcolor{Gainsboro!60}\text{Jordon \textit{et al.}~\cite{jordon2018pate}} & \text{2018} &  \text{DP}  & \text{GAN} & \text{\text{generating artificial biomedical data}}\\
                   \bottomrule
                \end{tabular}
                \label{tab:Crypto_dp}
            \end{table*}{}
            
        \section*{Differential Privacy}\label{sec:differential_privacy}
        
            One of the state-of-the-art concepts for eliminating and quantifying the chance of information leakage is \textit{differential privacy} \cite{ dwork2016calibrating, dwork2006our, nissim2017differential}.
            Differential privacy is a mathematical model that encapsulates the idea of injecting enough randomness or noise to sensitive data to camouflage the contribution of each single individual. This is achieved by inserting uncertainty into the learning process so that even a strong adversary with arbitrary auxiliary information about the data will still be uncertain in identifying any of the individuals in the dataset. This has become  standard in data protection and has been effectively deployed by Google \cite{erlingsson2014rappor} and Apple \cite{thakurta2017learning} as well as agencies such as the United States Census Bureau. Furthermore, it has drawn the attention of researchers in privacy-sensitive fields such as biomedicine and healthcare \cite{johnson2013privacy,  beaulieu2018privacy, fienberg2011privacy, uhlerop2013privacy, yu2014scalable,  han2018differential, tramer2015differential,vu2009differential, yu2014differentially,honkela2018efficient, han2019differential, simmons2016realizing, simmons2016enabling, wang2014differentially, wan2017controlling, al2017aftermath, fienberg2011privacy}.

            Differential privacy ensures that the model we  train does not overfit the sensitive data of a particular user. The model trained on a dataset containing information of a specific individual should be statistically indistinguishable from a model trained without the individual (Figure~\ref{fig:techniques}c). As an example, assume that a patient would like to give consent to his/her doctor to include his/her personal health record in a biomedical dataset to study the coordination between age and cardiovascular disease. Differential privacy provides a mathematical guarantee which captures the privacy risk associated with the patient's participation in the study and explains to what extent the analyst or the potential adversary can learn about that particular individual in the dataset. 
        
            Formally, a randomized algorithm (an algorithm that has randomness in its logic and whose output can vary even on a fixed input) $\mathit{A:D^{n}\xrightarrow{}Y}$ is ($\epsilon$, $\delta$)-differentially private if for all subsets $\mathit{y \subseteq Y}$ and for all adjacent datasets $D, D^\prime \in D^{n}$ that differ in at most one record the following inequality holds:
           \begin{equation*}
            Pr[A(D) \in y] \le e^{\epsilon}Pr[A(D') \in y]+\delta 
           \end{equation*} 
    
            Here, $\epsilon$ and $\delta$ are privacy loss parameters where lower values imply stronger privacy guarantees. $\delta$ is an exceedingly small value (e.g. $\num{e-5}$) indicating the probability of an uncontrolled breach, where the algorithm produces a specific output only in the presence of a specific individual and not otherwise. $\epsilon$ represents the worst case privacy breach in the absence of any such rare breach. If you assume $\delta = 0 $, you will have a pure ($\epsilon$)-differentially private algorithm, while if you consider  $\delta > 0 $ to approximate the case in which pure differential privacy is broken, you will have an approximate ($\epsilon$, $\delta$)-differentially private algorithm.
            
            Two important properties of differential privacy are composability \cite{kairouz2017composition} and resilience to post-processing. Composability means that combining multiple differentially private algorithms yields another differentially private algorithm. More precisely, if you combine $k$  ($\epsilon$, $\delta$)-differentially private algorithms, the composed algorithm is at least  ($k\epsilon$, $k\delta$)-differentially private. Differential privacy also assures resistance to post-processing theorem which states passing the output of an ($\epsilon$, $\delta$)-differentially private algorithm to any arbitrary randomized algorithm will still uphold the ($\epsilon$, $\delta$)-differential privacy guarantee. 
            
            The community efforts to ensure the privacy of sensitive genomic and biomedical data using differential privacy can be grouped into four categories according to the problem they address (Table~\ref{tab:Crypto_dp}): 
            \begin{enumerate}
                \item Approaches to querying  biomedical and genomics databases \cite{johnson2013privacy, wan2017controlling, al2017aftermath, cho2020privacy}. 
                \item  Statistical and AI modeling techniques in genomics and biomedicine \cite{ honkela2018efficient, vu2009differential, yu2014differentially, han2019differential, simmons2016realizing, simmons2016enabling}.
                \item Data release, i.e., releasing summary statistics of a GWAS such as $p$-values and $\chi^2$ contingency tables  \cite{fienberg2011privacy, uhlerop2013privacy,yu2014scalable, wang2014differentially}.
                \item Training privacy-preserving generative models \cite{abay2018privacy, beaulieu2019privacy, jordon2018pate}.
            \end{enumerate}
            
            Studies in the first category proposed solutions to reduce the privacy risks of  genomics databases such as GWAS databases and genomics beacon service \cite{fiume2019federated}. The Beacon Network \cite{beacon} is an online web service developed by the Global  Alliance for Genomics and Health (GA4GH) through which the users can query the data provided by owners or research institutes, ask about the presence of a genetic variant in the database, and get a YES/NO as response. Studies have shown that an attacker can detect membership in the Beacon or GWAS by querying these databases multiple times and asking different questions \cite{shringarpure2015privacy, raisaro2018protecting, hardt2012simple, raisaro2018m}. Very recently, Cho \textit{et al.} \cite{cho2020privacy} proposed a theoretical differential privacy mechanism to maximize the utility of count query in biomedical systems while guaranteeing data privacy. Johnson \textit{et al.} \cite{johnson2013privacy} developed a differentially private query-answering framework. With this framework an analyst can retrieve statistical properties such as the correlation between SNPs and get an almost accurate answer while the GWAS dataset is protected against privacy risks. In anoher study, Aziz \textit{et al.} \cite{al2017aftermath} proposed two algorithms to make the Beacon’s response inaccurate by controlling a bias variable. These algorithms decide when to answer the query correctly/incorrectly according to specific conditions in the bias variable so that it gets harder for the attacker to succeed.
            
            Some of the efforts in the second category addressed the privacy concerns in  GWAS by introducing differentially private logistic regression to identify associations between SNPs and diseases \cite{han2019differential} or associations among multiple SNPs \cite{yu2014differentially}. Honkela \textit{et al.} \cite{honkela2018efficient} improve drug sensitivity prediction by effectively employing differential privacy for Bayesian linear regression. Moreover, Simmons \textit{et al.} \cite{simmons2016enabling} presented a differentially private EIGENSTRAT (PrivSTRAT) \cite{price2006principal} and linear mixed model (PrivLMM) \cite{yang2014advantages} to correct for population stratification.  In another paper, Simmons \textit{et al.} \cite{simmons2016realizing} tackled the problem of finding significant SNPs  by modeling it as an optimization problem. Solving this problem provides a differentially private estimate of the neighbor distance for all SNPs so that high scoring SNPs can be found. 
            
            The third category focused on releasing summary statistics such as $p$-values, $\chi^2$ contingency tables, and minor allele frequencies in a differentially private fashion. The common approach in these studies is to add Laplacian noise to the true value of the statistics, so that sharing the perturbed statistics preserves privacy of the individuals. They vary in the sensitivity of the algorithms (that is, the maximum change on the output of an algorithm in presence or absence of a specific data point) and hence require different amounts of injected noise \cite{fienberg2011privacy, uhlerop2013privacy, wang2014differentially}.
            
            The forth category proposed novel privacy-protecting methods to generate synthetic healthcare data leveraging differentially private generative models (Figure ~\ref{fig:GM}). Deep generative models, such as generative adversarial networks (GANs) \cite{gan}, can be trained on sensitive genomics and biomedical data to capture its properties and generate artificial data with similar characteristics as the original data. 
            \noindent   
            \begin{minipage}{\linewidth}
                \medskip
                 \centering
                 \includegraphics[width=\textwidth]{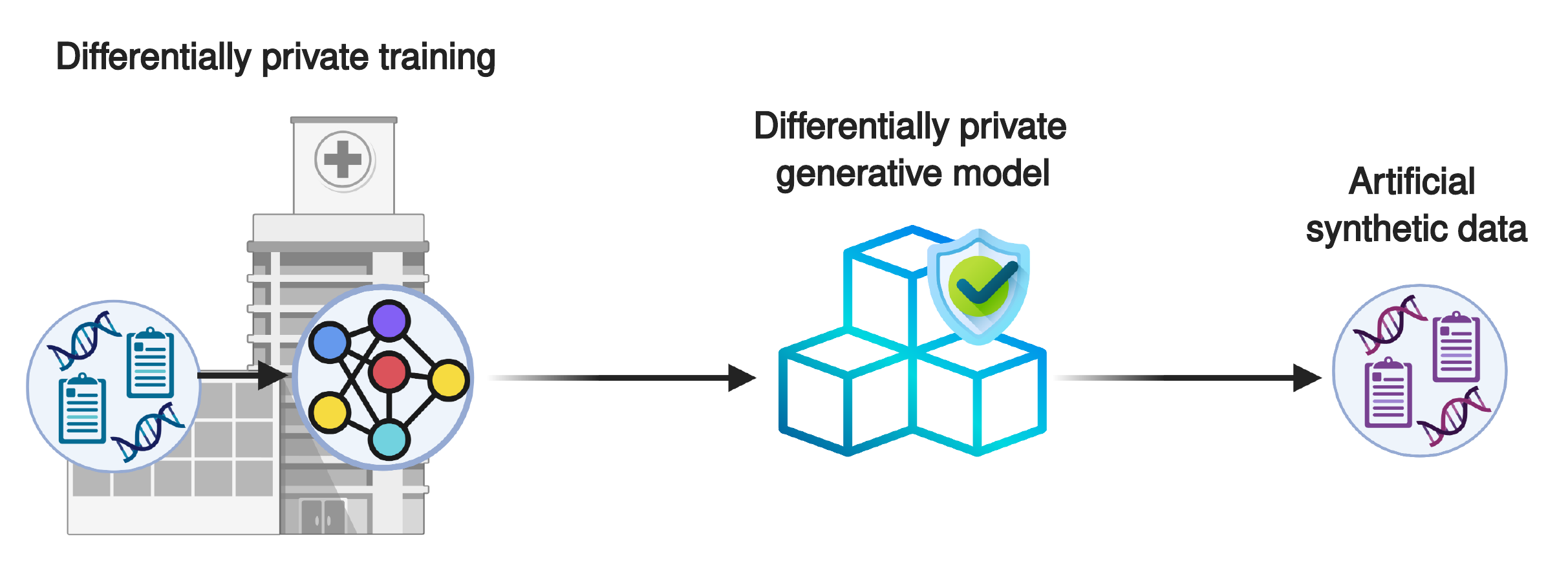}
                 \captionof{figure}{\textbf{Differentially private deep generative models:} The sensitive data holder (e.g. health institutes) train a differentially private generative model locally and share just the trained data generator with the outside world (e.g. researchers). The shared data generator can then be used to produce artificial data with the same characteristics as the sensitive data.}
                 \label{fig:GM}
                \medskip
            \end{minipage}
        
            Abay \textit{et al.} \cite{abay2018privacy} presented a differentially private deep generative model, DP-SYN, a generative autoencoder that splits the input data into multiple partitions, then learns and simulates the representation of each partition while maintaining the privacy of input data. They assessed the performance of DP-SYN on sensitive datasets of breast cancer and diabetes. Beaulieu \textit{et al.} \cite{beaulieu2019privacy} trained an auxiliary classifier GAN (AC-GAN) in a differentially private manner to simulate the participants of the SPRINT trial (Systolic Blood Pressure Trial), so that the clinical data can be shared while respecting participants' privacy. In another approach, Jordon \textit{et al.} \cite{jordon2018pate}  introduced a differentially private GAN, PATE-GAN, and evaluated the quality of synthetic data on Meta-Analysis Global Group in Chronic Heart Failure (MAGGIC) and the United Network for Organ Transplantation (UNOS) datasets.
            
            Despite the aforementioned achievements in adopting differential privacy in the field, several challenges remain to be addressed. Although differential privacy involves less network communication, memory usage and time complexity compared to cryptographic techniques, it still struggles with giving highly accurate results within a reasonable privacy budget, namely, intended $\epsilon$ and $\delta$, on large-scale datasets such as genomics datasets \cite{wang2017genome, aziz2019privacy}. In more details, since genomic datasets are huge, the sensitivity of the applied algorithms on these datasets is large. Hence, the amount of distortion required for anonymization increases significantly, sometimes to the extent that the results will not be meaningful anymore \cite{KiesebergEtAl:2014:protecting}. Therefore, to make differential privacy more practical in the field, balancing a trade off between privacy and utility demands more attention than it has received \cite{vu2009differential, han2018differential, tramer2015differential, wang2014differentially}.
            
        \section*{Federated Learning}\label{sec:federated_learning}
        Federated learning \cite{mcmahan2016federated} is a type of distributed learning where multiple clients (e.g. hospitals) collaboratively learn a model under the coordination of a central server while preserving the privacy of their data. Instead of sharing its private data with the server or the other clients, each client extracts knowledge (that is, model parameters) from its data and transfers it to the server for aggregation (Figure~\ref{fig:techniques}d).

        Federated learning is an iterative process in which each iteration consists of the following steps \cite{kairouz2019advancesfederated}:
        \begin{enumerate}
            \item The server chooses a set of clients to participate in the current iteration of the model.
            \item The selected clients obtain the current model from the server.
            \item Each selected client computes the local parameters using the current model and its private data (e.g., runs gradient descent algorithm initialized by the current model on its local data to obtain the local gradient updates).
            \item The server collects the local parameters from the selected clients and aggregates them to update the current model.
        \end{enumerate}
   
        The data of the clients can be considered as a table, where rows represent samples (e.g., individuals) and columns represent features or labels (e.g., age, case vs.\ control). We refer to the set of samples, features, and labels of the data as \textit{sample space}, \textit{feature space}, and \textit{label space}, respectively. Federated learning can be categorized into three types based on the distribution characteristics of the clients' data:
        \begin{itemize}
            \item \textbf{Horizontal (sample-based) federated learning} ~\cite{yang2019federatedconcepts}: Data from different clients shares similar feature space but is very different in sample space. As an example, consider two hospitals in two different cities which collected similar information such as age, gender, and blood pressure about the individuals. In this case, the feature spaces are similar; but because the individuals who participated in the hospitals' data collections are from different cities, their intersection is most probably very small, and the sample spaces are hence very different.
            \item \textbf{Vertical (feature-based) federated learning} ~\cite{yang2019federatedconcepts} : Clients' data is similar in sample space but very different in feature space. For example, two hospitals with different expertise in the same city might collect different information (different feature space) from almost the same individuals (similar sample space).
            \item \textbf{Hybrid federated learning}: Both feature space and sample space are different in the data from the clients. For example, consider a medical center with expertise in brain image analysis located in New York and a research center with expertise in protein research based in Berlin. Their data is completely different (image vs. protein data) and disjoint groups of individuals participated in the data collection of each center.
        \end{itemize}
        
        To illustrate  the concept of federated learning, consider a scenario with two hospitals $A$ and $B$. $A$ and $B$ possess lists $X$ and $Y$, containing the age of their cancer patients, respectively. A simple federated mean algorithm to compute the average age of cancer patients in both hospitals without revealing the real values of $X$ and $Y$ works as follows: For the sake of brevity, we assume that both hospitals are selected in the first step and that the current global model parameters (average age) in the second step are zero (see federated learning steps). 
        \begin{itemize}
            \item Hospital $A$ computes the average age ($M_X$) and number of its cancer patients ($N_X$). Hospital $B$ does the same, resulting in $M_Y$, $N_Y$. Here, $X$ and $Y$ are private data while $M_X$, $N_X$, $M_Y$, $N_Y$ are the parameters extracted from the private data.
            \item The server obtains the values of local model parameters from the hospitals and computes the global mean as follows:
            \begin{equation*}
                M_G = \frac{M_X \times N_X + M_Y \times N_Y }{N_X + N_Y }
            \end{equation*}
        \end{itemize}
        
        The emerging demand for federated learning gave rise to a wealth of both simulation \cite{TFF, ryffel2018generic} and production-oriented \cite{FATE, PaddleFL} open source frameworks. Additionally, there are AI platforms whose goal is to apply federated learning to real-world healthcare settings \cite{FC, ClaraFL}. In the following, we survey studies on federated AI techniques in biomedicine and healthcare (Table~\ref{tab:FL_hybrid}). Recent studies in this regard mainly focused on horizontal federated learning and there are a few vertical or hybrid federated learning algorithms applicable to genomic and biomedical data. 
        
        A number of the studies provided solutions for the lack of sufficient data due to the the privacy challenges in the medical imaging domain \cite{sheller2018multi,vepakomma2018split, vepakomma2019reducing,poirot2019split, balachandar2020accounting, chang2018distributed, ClaraFL}. For instance, Sheller \textit{et al.} developed a supervised DNN in a federated way for semantic segmentation of brain Gliomas from magnetic  resonance  imaging scans \cite{sheller2018multi}. Chang \textit{et al.} \cite{chang2018distributed} simulated a distributed DNN in which multiple participants collaboratively update model weights using training heuristics such as single weight transfer and cyclical weight transfer (CWT). They evaluated this distributed model using image classification tasks on medical image datasets such as mammography and retinal fundus image collections, which were evenly distributed among the participants. Balachandar \textit{et al.} \cite{balachandar2020accounting} optimized CWT for cases where the datasets are unevenly distributed across participants. They assessed their optimization methods on simulated diabetic retinopathy detection and chest radiograph classification.
        
        Federated Cox regression, linear regression, logistic regression as well as chi-square test have been developed for sensitive biomedical data that is vertically or horizontally distributed \cite{ dai2020verticox, lu2015webdisco, wu2012g, wang2013expectation, li2016vertical, Nasirigerdeh2020splink}. VERTICOX~\cite{dai2020verticox} is a vertical federated Cox regression model for survival analysis, which employs the alternating direction method of multiplier (ADMM) framework~\cite{admm} and is evaluated on acquired immunodeficiency syndrome (AIDS) and breast cancer survival datasets. Similarly, WebDISCO \cite{lu2015webdisco} presents a federated Cox regression model but for horizontally distributed survival data. The grid binary logistic regression (GLORE) \cite{wu2012g} and the expectation propagation logistic regression (EXPLORER) \cite{wang2013expectation} implemented a horizontally federated logistic regression for GWAS data. Unlike GLORE, EXPLORER supports asynchronous communication and online learning functionality so that the system can continue collaborating in case a participant is absent or if communication is interrupted. Li \emph{et al.} presented  VERTIGO \cite{li2016vertical}, a vertical grid logistic regression algorithm designed for vertically distributed biological datasets such as breast cancer genome and myocardial infarction data. Nasirigerdeh \textit{et al.} ~\cite{Nasirigerdeh2020splink} developed a horizontally federated tool set for GWAS, called \textit{sPLINK}, which supports chi-square test, linear regression, and logistic regression. Notably, federated results from \textit{sPLINK} on distributed datasets are the same as those from aggregated analysis conducted with \textit{PLINK}~\cite{purcell2007plink}. Moreover, they showed that \textit{sPLINK} is robust against heterogeneous (imbalanced) data distributions across clients and does not lose its accuracy in such scenarios.
        
        There are also studies that combine federated learning with other traditional AI modeling techniques such as ensemble learning, support vector machines (SVMs) and principle component analysis (PCA) \cite{brisimi2018federated,huang2018loadaboost, liu2018fadl, chen2019fedhealth, silva2019federated}. Brisimi \textit{et al.} \cite{brisimi2018federated} presented a federated soft-margin support vector machine (sSVM) for distributed electronic health records. Huang \textit{et al.} \cite{huang2018loadaboost} introduced LoAdaBoost, a federated adaptive boosting method for learning biomedical data such as intensive care unit data from distinct hospitals \cite{pollard2018eicu} while Liu \textit{et al.} \cite{liu2018fadl} trained a federated autonomous deep learner to this end. There have also been a couple of attempts at incorporating federated learning into multi-task learning and transfer learning in general \cite{smith2017federated, corinzia2019variational, liu2018secure}. However, to the best of our knowledge, FedHealth \cite{chen2019fedhealth} is the only federated transfer learning framework specifically designed for healthcare applications. It enables users to train personalized models for their wearable healthcare devices by aggregating the data from different organizations without compromising privacy. 
        
        \begin{table*}[ht]
            \centering
            \caption{Summary of \textbf{federated learning} (FL) and \textbf{hybrid} approaches in genomics and biomedicine}
            \medskip
            \begin{tabular}{lSlll}\toprule
                \textbf{Authors}  & \textbf{Year}  & \textbf{Privacy Technique}  & \textbf{Model} & \textbf{Application}  \\ 
                \midrule
                
                \rowcolor{Gainsboro!60}\text{Sheller \textit{et al.}~\cite{sheller2018multi}}  & \text{2018} & \text{FL} & \text{DNN} & \text{medical image segmentation} \\
                
                \text{\makecell[l]{Chang \textit{et al.}~\cite{chang2018distributed} \\ Balachandar \textit{et al.}~\cite{balachandar2020accounting}}}  & \text{\makecell[l]{2018 \\ 2020}} & \text{FL} & \text{\makecell[l]{single weight transfer \\ cyclical weight transfer}} & \text{medical image classification}\\
                
                \rowcolor{Gainsboro!60}\text{Nasirigerdeh \textit{et al.}~\cite{Nasirigerdeh2020splink}}  & \text{2020}  & \text{FL}& \text{\makecell[l]{linear regression\\ chi-square\\ logistic regression}} & \text{genetic associations} \\

                \text{\makecell[l]{Wu \textit{et al.}~\cite{wu2012g} \\ Wang \textit{et al.}~\cite{wang2013expectation} \\ Li \textit{et al.}~\cite{li2016vertical} }}  & \text{\makecell[l]{2012 \\ 2013 \\ 2016}}  & \text{FL}& \text{logistic regression} & \text{genetic associations} \\
                
                \rowcolor{Gainsboro!60}\text{\makecell[l]{ Dai \textit{et al.}~\cite{dai2020verticox} \\ Lu \textit{et al.}~\cite{lu2015webdisco} }}  & \text{\makecell[l]{2020 \\ 2015}} & \text{FL} & \text{cox regression} & \text{survival analysis} \\
                
                \text{Brisimi \textit{et al.}~\cite{brisimi2018federated}}  & \text{2018} & \text{FL} & \text{support vector machine} & \text{classifying electrical health records}\\
                
                \rowcolor{Gainsboro!60}\text{Huang \textit{et al.}~\cite{huang2018loadaboost}}  & \text{2018}& \text{FL}  & \text{adaptive boosting ensemble} & \text{classifying medical data}\\
                
                \text{Liu \textit{et al.}~\cite{liu2018fadl}}  & \text{2018} & \text{FL} & \text{autonomous deep learning} & \text{classifying medical data}\\
                
                \rowcolor{Gainsboro!60}\text{Chen \textit{et al.}~\cite{chen2019fedhealth}}  & \text{2019}  & \text{FL}  & \text{transfer learning} & \text{training wearable healthcare devices}\\
                
                \midrule
                \text{Li \textit{et al.}~\cite{li2019privacy}}  & \text{2019} & \text{FL+DP}  & \text{DNN} & \text{medical image segmentation}\\
                
                \rowcolor{Gainsboro!60}\text{Li \textit{et al.}~\cite{li2020multi}}  & \text{2020} & \text{FL+DP}  & \text{domain adoption} & \text{medical image pattern recognition} \\
                
                \text{Choudhury \textit{et al.}~\cite{choudhury2019differential}} &\text{2019} & \text{FL+DP}  & \text{\makecell[l]{perceptron neural network \\ support vector machine \\ logistic regression }} & \text{classifying electronic health records} \\
                
                \rowcolor{Gainsboro!60}\text{Constable \textit{et al.}~\cite{constable2015privacy}}  & \text{2015} & \text{FL+SMPC }  & \text{\makecell[l]{statistical analysis \\ (e.g. $\chi^2$ statistics)}} & \text{genetic associations}\\
                
                \text{Lee \textit{et al.}~\cite{lee2018privacy}}  & \text{2018} & \text{FL+HE }  & \text{context-specific hashing} & \text{learning patient similarity}\\
                
                \rowcolor{Gainsboro!60}\text{Kim \textit{et al.}~\cite{kim2019secure}} & \text{2019} & \text{FL+DP+HE }  & \text{logistic regression} & \text{classifying medical data}\\
                
                \bottomrule
            \end{tabular}
            \label{tab:FL_hybrid}
        \end{table*}{}
        
        One of the major challenges for adopting federated learning in large scale genomics and biomedical applications is the significant network communication overhead, especially for complex AI models such as DNNs that contain millions of model parameters and require thousands of iterations to converge. A rich body of literature exists to tackle this challenge, known as communication-efficient federated learning \cite{gupta2015gradquant,aji2017gradsparse,mcmahan2016communication,tang2020communication}.
        
        Another challenge in federated learning is the possible accuracy loss from the aggregation process if the data distribution across the clients is heterogeneous (i.e. not independent and identically distributed (IID)). More specifically, federated learning can deal with non-IID data while preserving the model accuracy if the learning model is simple such as ordinary least squares (OLS) linear regression (\textit{sPLINK} \cite{Nasirigerdeh2020splink}). However, when it comes to learning complex models such as DNNs, the global model might not converge on non-IID data across the clients. \textit{Zhao et al.}~\cite{zhao2018non-iid1} showed that simple averaging of the model parameters in the server significantly diminishes the accuracy of a convolutional neural network model in highly skewed non-IID settings. Developing the aggregation strategies which are robust against non-IID scenarios is still an open and interesting problem in federated learning.
        
        Finally, federated learning is based on the assumption that the centralized server is honest and not compromised, which is not necessarily the case in real applications. To relax this assumption, differential privacy or cryptographic techniques can be leveraged in federated learning, which is covered in the next section. For further reading on future directions of federated learning in general, we refer the reader to comprehensive surveys \cite{li2019federated, kairouz2019advancesfederated, rieke2020future}.
       
          
        \section*{Hybrid Privacy-preserving Techniques}\label{sec:hybrid}
        
          The hybrid techniques combine federated learning with the other paradigms (cryptographic techniques and differential privacy) to enhance privacy or provide privacy guarantees (Table~\ref{tab:FL_hybrid}). Federated learning preserves privacy to some extent because it does not require the health institutes to share the patients' data with the central server. However, the model parameters that participants share with the server might be  abused to reveal the underlying private data if the coordinator is compromised \cite{melis2019exploiting}. To handle this issue, the participants can leverage differential privacy and add noise to the model parameters before sending them to the server (FL+DP) \cite{geyer2017differentially, li2019privacy, li2020multi,  truex2019hybrid, wei2020federated} or they employ HE (FL+HE)\cite{hardy2017private, zhang2020batchcrypt, sadat2018safety}, SMPC (FL+SMPC) or both DP and HE (FL+DP+HE) \cite{kim2019secure, raisaro2018m, froelicher2017unlynx} to securely share the parameters with the server \cite{lee2018privacy, constable2015privacy}.
          
          In the genomic and biomedical field, several hybrid approaches have been presented recently. Li \textit{et al.} \cite{li2019privacy} presented a federated deep learning framework for magnetic resonance brain image segmentation in which the client side provides differential privacy guarantees on selecting and sharing the local gradient weights with the server for imbalanced data. A recent study \cite{li2020multi} extracted neural patterns from brain functional magnetic resonance images by developing a privacy-preserving pipeline that analyzes image data of patients having different psychiatric disorders using federated domain adaption methods. Choudhury \textit{et al.} \cite{choudhury2019differential} developed a federated differential privacy mechanism for gradient-based classification on electronic health records. 
                     
          There are also some studies that incorporate federate learning with cryptographic techniques. For instance, Constable \textit{et al.} \cite{constable2015privacy} implemented a privacy-protecting structure for federated statistical analysis such as $\chi^2$ statistics on GWAS while maintaining privacy using SMPC. In a slightly different approach, Lee \textit{et al.} \cite{lee2018privacy} presented a privacy-preserving platform for learning patient similarity in multiple hospitals using a context-specific hashing approach which employs homomorphic encryption to limit the privacy leakage. Moreover, Kim \textit{et al.} \cite{kim2019secure} presented a privacy-preserving federated logistic regression algorithm for horizontally distributed diabetes and intensive care unit datasets. In this approach, the logistic regression ensures privacy by making the aggregated weights differentially private and encrypting the local weights using homomorphic encryption. 
    
          Incorporating HE, SMPC,  and differential privacy into federated learning brings about enhanced privacy but it combines the limitations of the approaches, too. FL+HE puts much more computational overhead on the server, since it requires to perform aggregation on the encrypted model parameters from the clients. The network communication overhead is exacerbated in FL+SMPC,  because clients need to securely share the model parameters with multiple computing parties instead of one. FL+DP might result in inaccurate models because of adding noise to the model parameters in the clients.
          
        \begin{table*}[!htb]
            \centering
            \caption{Comparison among the privacy-preserving techniques including homomorphic encryption (HE), secure multiparty computation (SMPC), federated learning (FL), differential privacy (DP) and the hybrid approaches (FL+DP, FL+HE and FL+SMPC); The generic ranking (lowest =1 to highest = 6) is used for comparison purposes such that having a higher score for a criteria, represents better performance.} 
            \medskip
            \label{tab:comparison} 
            \begin{tabular}{cccccccc}\toprule
                
                \textbf{ }  & \textbf{HE}  & \textbf{SMPC}  & \textbf{DP}  & \textbf{FL}  & \textbf{FL+DP}  & \textbf{FL+HE}  & \textbf{FL+SMPC} \\ 
                \midrule

                \rowcolor{Gainsboro!60}
                \text{Accuracy}  & \text{2}  & \text{6} & \text{1}  & \text{5} & \text{3}  & \text{4} & \text{5}\\

                \text{Computational efficiency}  & \text{1}  & \text{2} & \text{NA}  & \text{6} & \text{5}  & \text{3} & \text{4}\\
                
                \rowcolor{Gainsboro!60}
                \text{Network communication efficiency}  & \text{5}  & \text{4} & \text{NA}  & \text{3} & \text{3}  & \text{2} & \text{1} \\

                \text{Privacy of exchanged traffic}  & \text{4}  & \text{3} & \text{NA}  & \text{1} & \text{2}  & \text{4} & \text{3}\\
                
                \rowcolor{Gainsboro!60}
                \text{Exchanging low sensitive traffic}  & \text{\xmark}  & \text{\xmark} & \text{NA}  & \text{\cmark}  & \text{\cmark} & \text{\cmark} & \text{\cmark} \\
                 
                \text{Privacy guarantee}  & \text{\xmark}  & \text{\xmark} & \text{\cmark}  & \text{\xmark} & \text{\cmark}  & \text{\xmark} & \text{\xmark} \\
            \bottomrule
            \end{tabular}
            \end{table*}{}

        \section*{Comparison}\label{sec:comparison}
\begin{figure*}[!htb]
                \begin{minipage}{.26\textwidth}
                \centering       \includegraphics[width=1\textwidth]{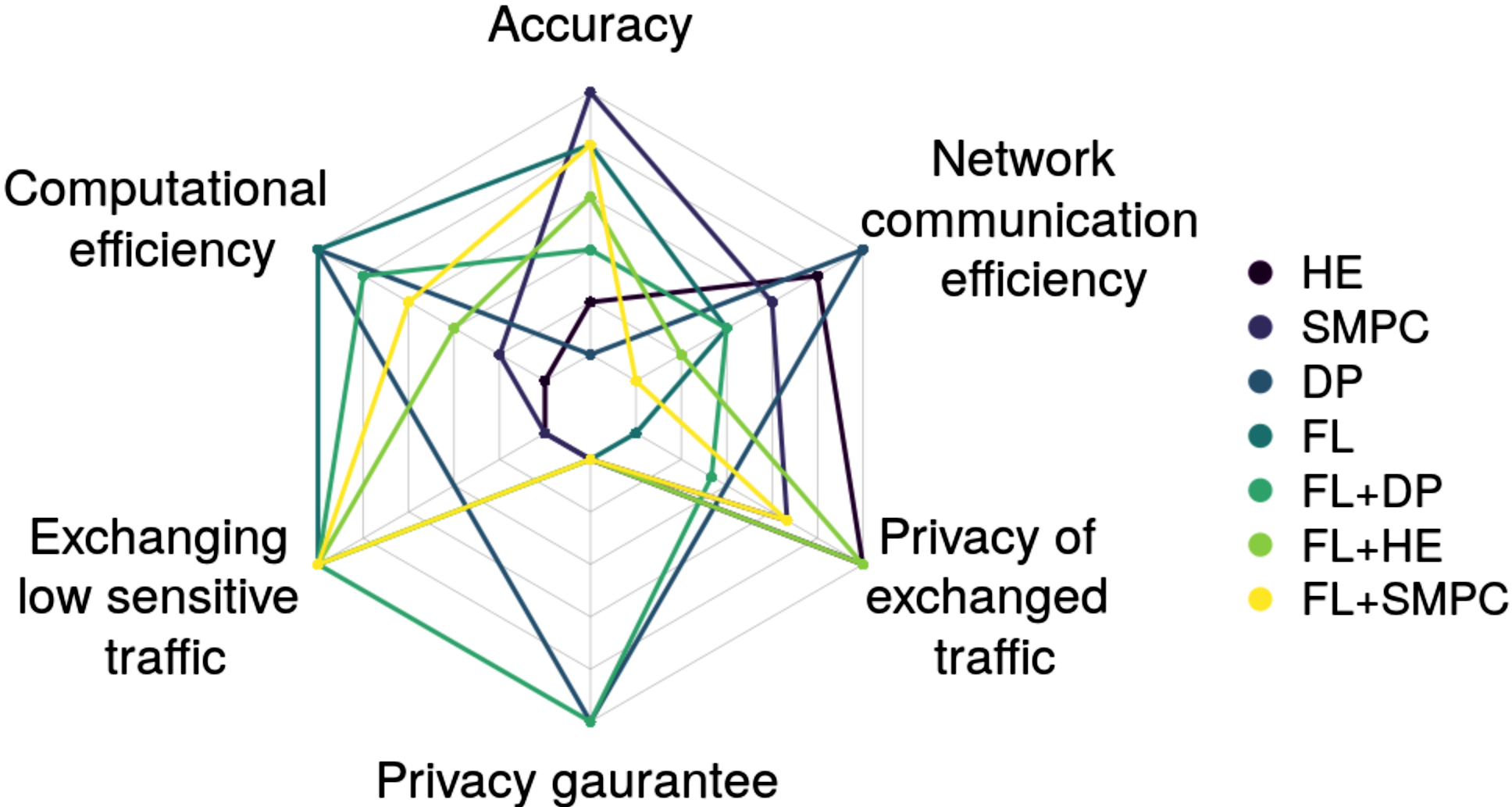}
                    \subcaption{) All }  
                \end{minipage}
                \hfill
                \begin{minipage}{.22\textwidth}
                \centering                   \includegraphics[width=1.0\textwidth]{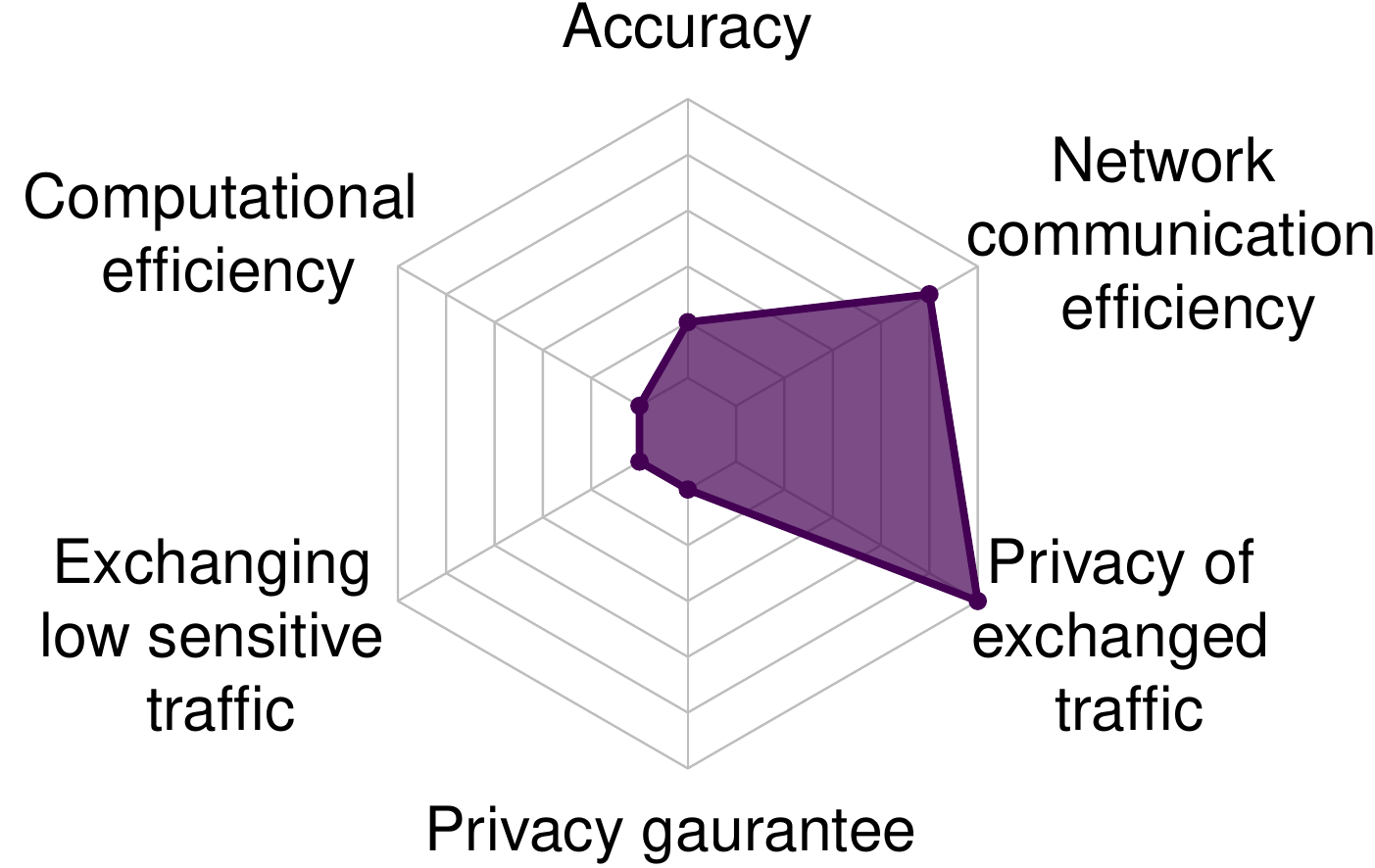}
                    \subcaption{) HE }  
                \end{minipage}
                \hfill
                \begin{minipage}{.22\textwidth}
                \centering                 \includegraphics[width=1.0\textwidth]{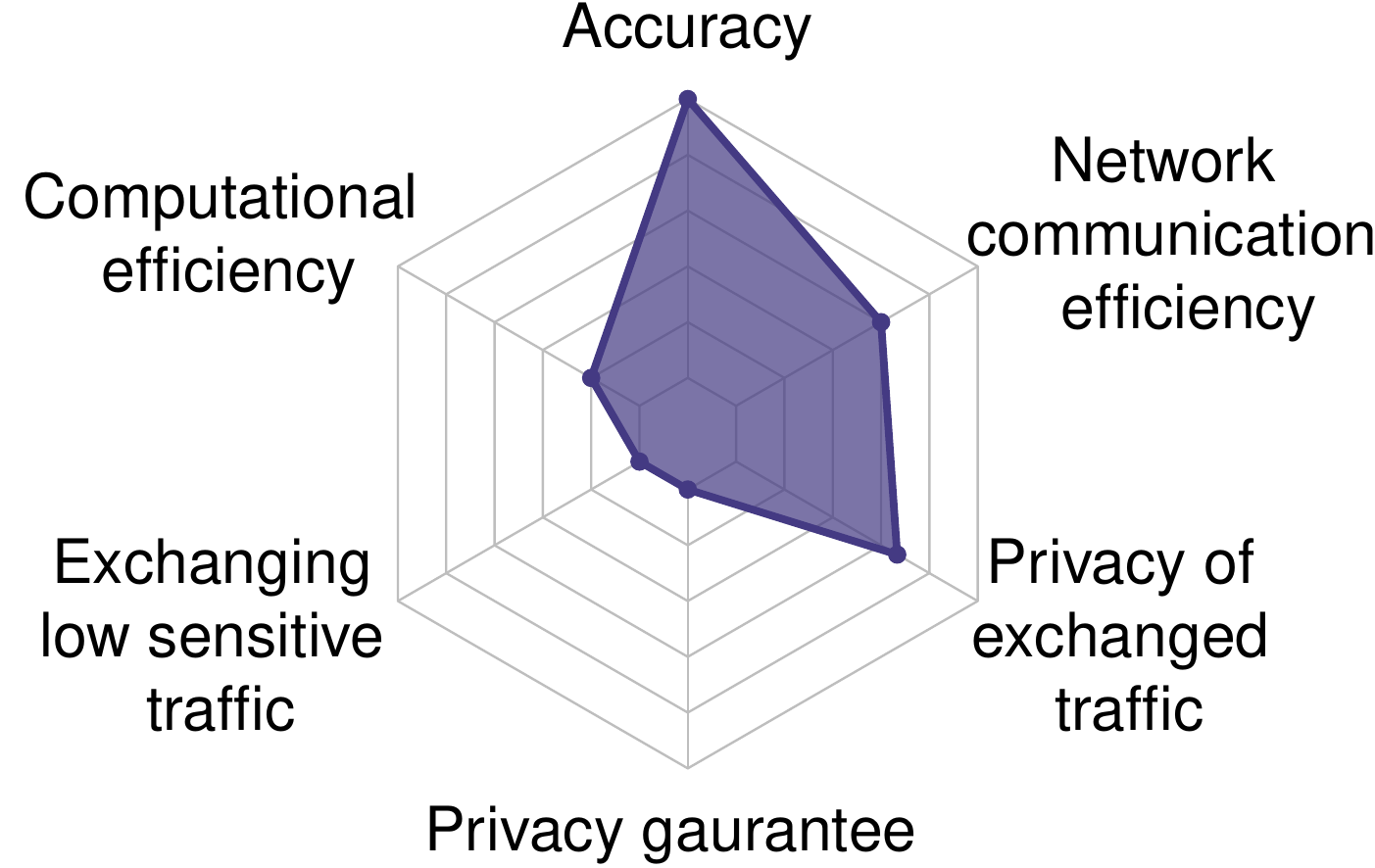}
                    \subcaption{ ) SMPC }  
                \end{minipage}
                \hfill
                \begin{minipage}{0.22\textwidth}
                \centering                  \includegraphics[width=1.0\textwidth]{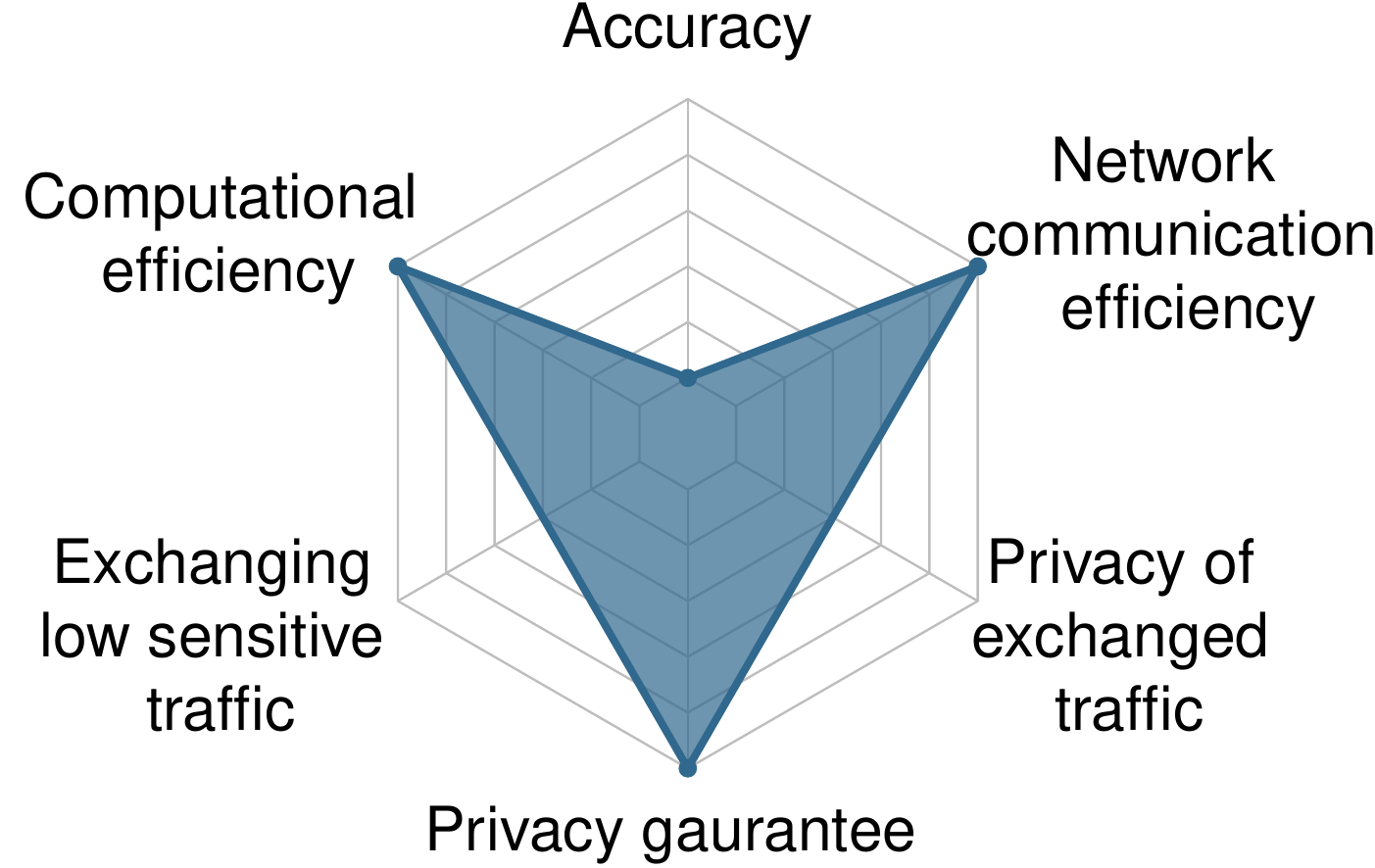}
                    \subcaption{) DP }  
                \end{minipage}
                
                \leavevmode \newline
                \begin{minipage}{.22\textwidth}
                \bigskip
                \centering \includegraphics[width=1.0\textwidth]{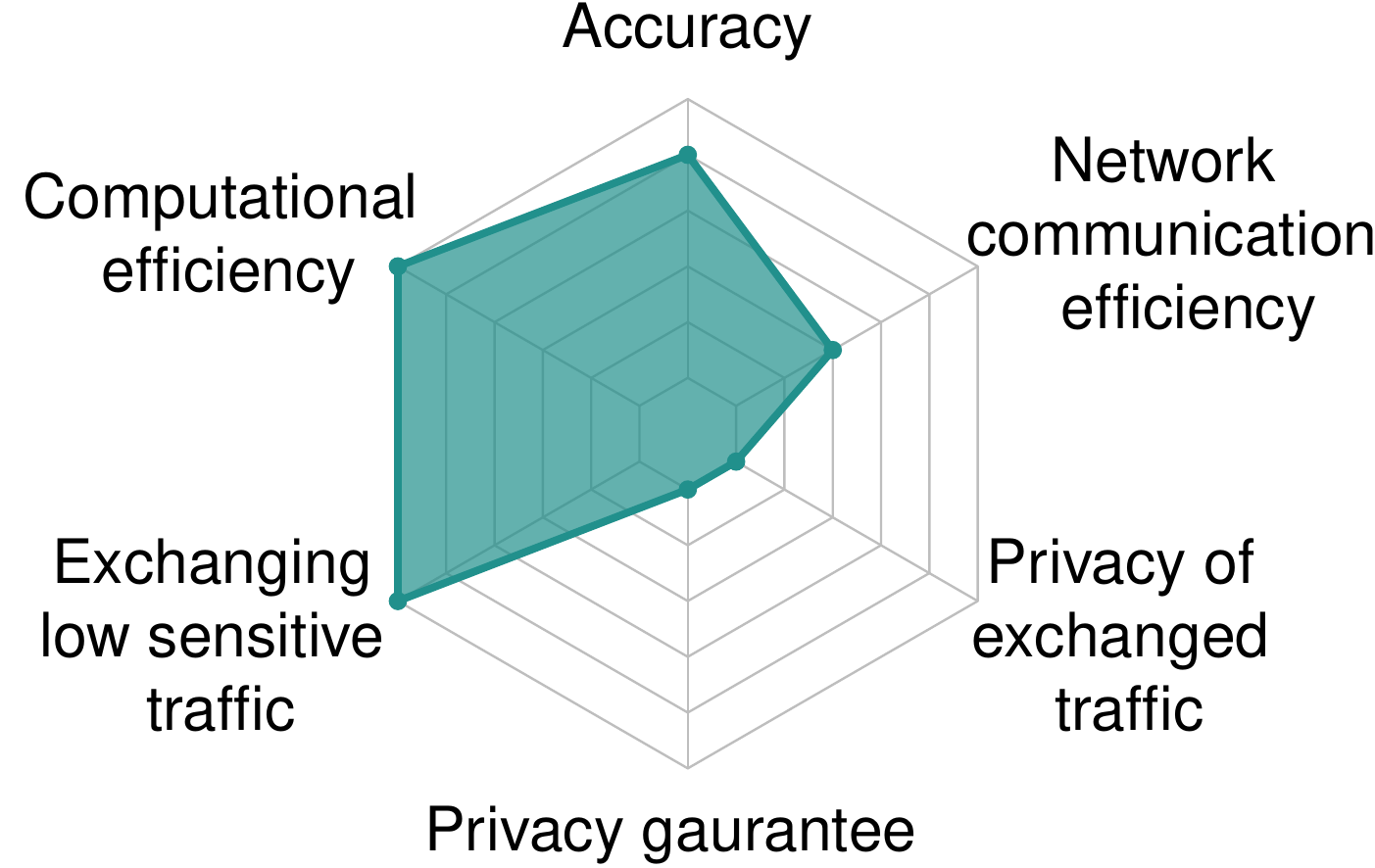}
                    \subcaption{) FL }  
                \end{minipage}
                \hfill
                \begin{minipage}{.22\textwidth}
                    \bigskip                    \includegraphics[width=1.0\textwidth]{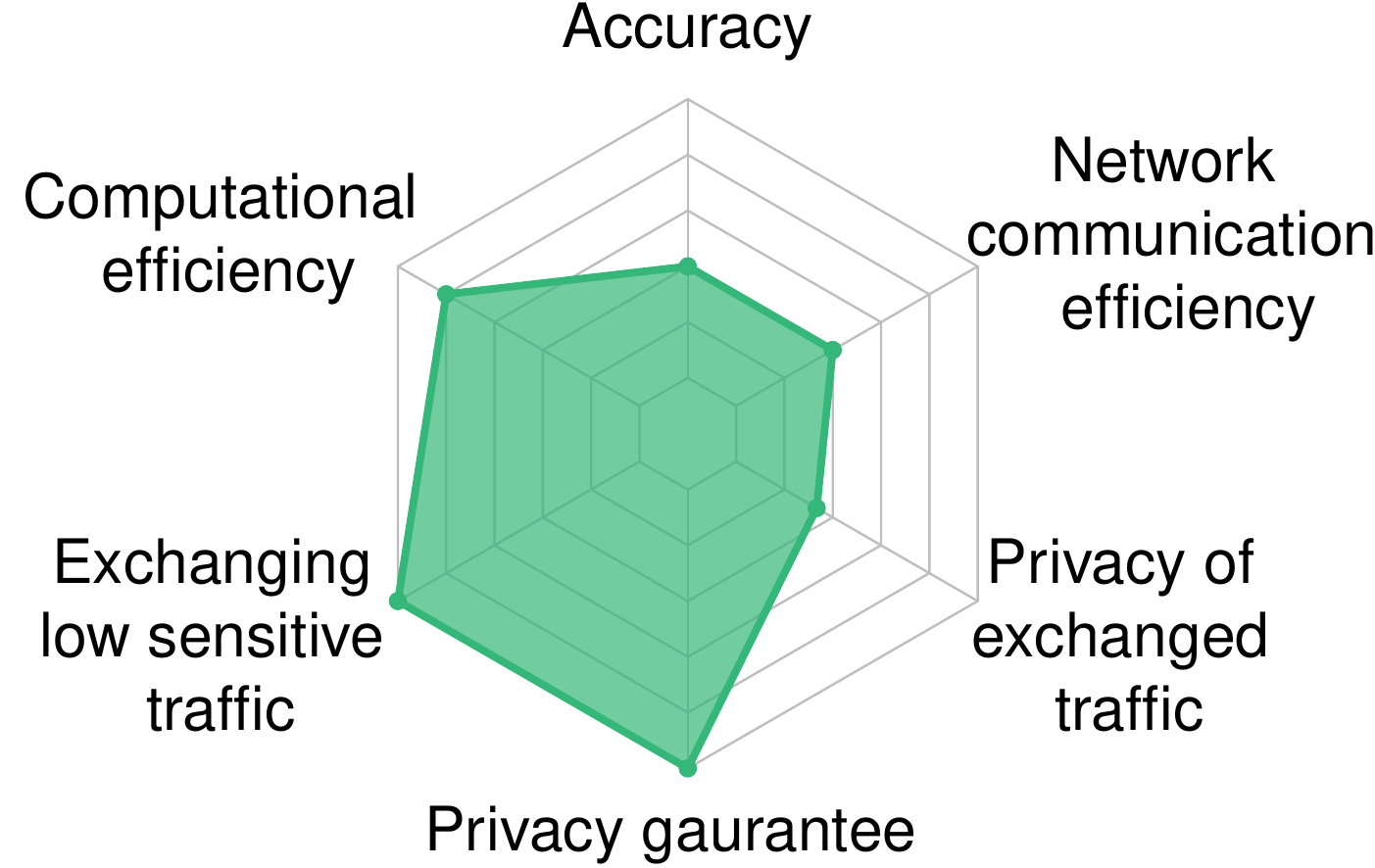}
                    \subcaption{) FL+DP  }  
                \end{minipage}
                \hfill
                \begin{minipage}{.22\textwidth}
                    \bigskip \includegraphics[width=1.0\textwidth]{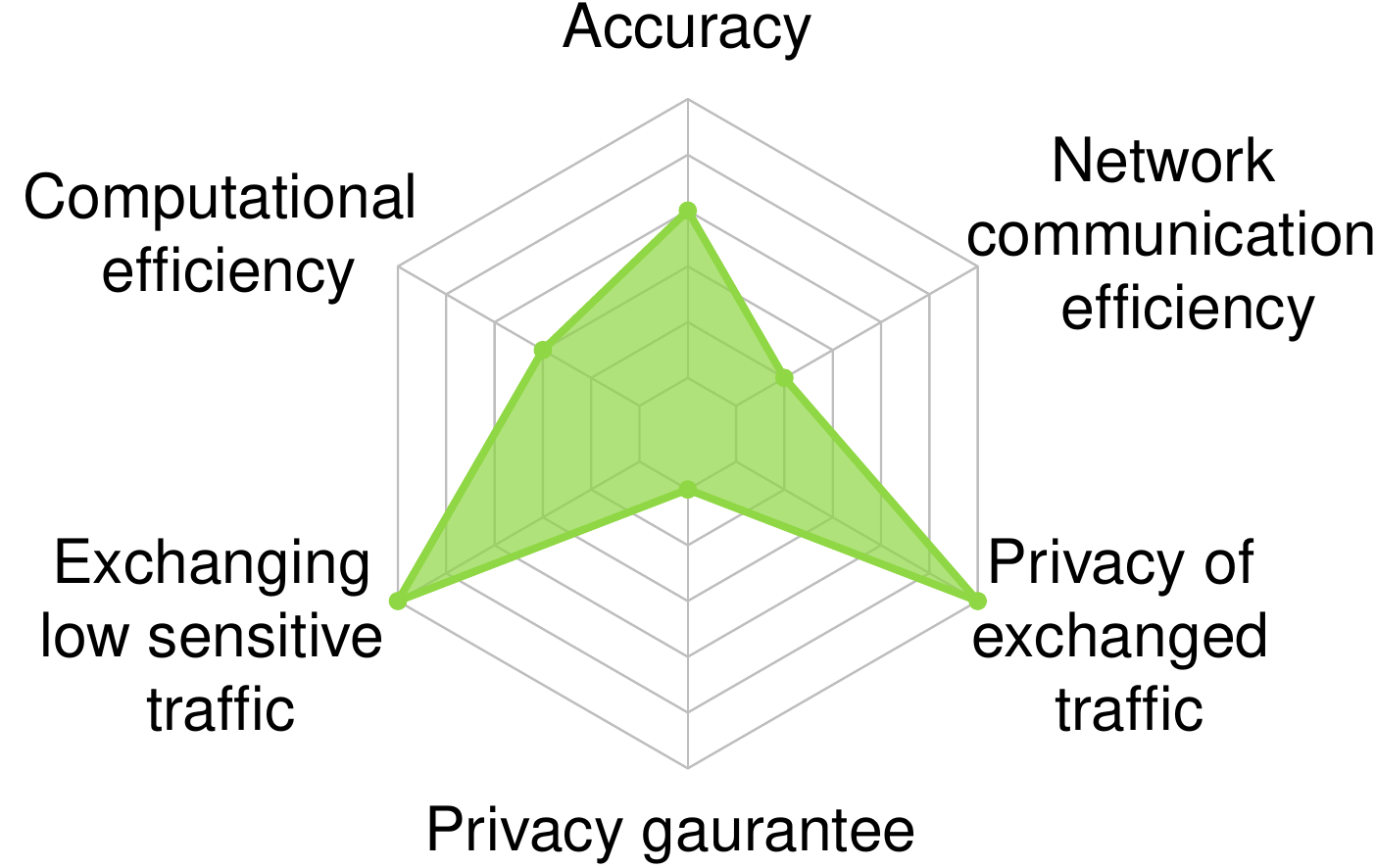}
                    \subcaption{) FL+HE}  
                \end{minipage}
                \hfill
                \begin{minipage}{0.22\textwidth}
                    \bigskip                    \includegraphics[width=1.0\textwidth]{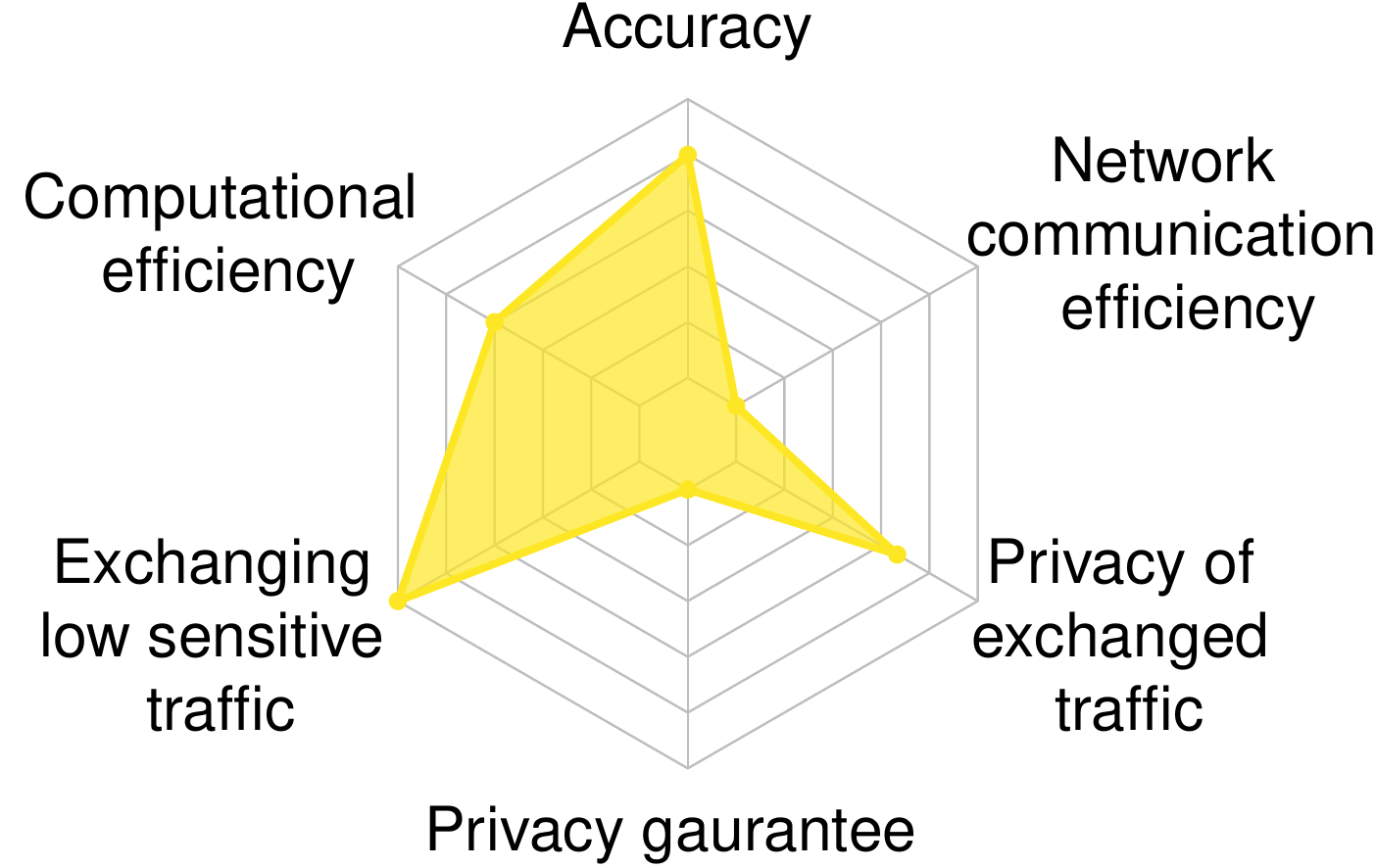}
                    \subcaption{) FL+SMPC}  
                \end{minipage}
                
                \caption{Comparison radar plots for all (a) and each of (b-h) the privacy preserving approaches including homomorphic encryption (HE), Secure multiparty computation (SMPC), differential privacy (DP), federated learning (FL) and hybrid techniques (FL+DP, FL+HE and FL+SMPC)}
                \label{fig:comparison}
            \end{figure*}

             We compare the privacy-preserving techniques (HE, SMPC, differential privacy, federated learning, and the hybrid approaches) using various performance and privacy criteria such as \textit{computational/communication efficiency}, \textit{accuracy}, \textit{privacy guarantee}, \textit{exchanging sensitive traffic through network} and \textit{privacy of exchanged traffic} (Table \ref{tab:comparison} and Figure \ref{fig:comparison}). We employ a generic ranking (lowest =1 to highest = 6) \cite{aziz2019privacy} for all comparison criteria except for \textit{privacy guarantee} and \textit{exchanging sensitive traffic through network}, which are binary criteria. This comparison is made under the assumption of applying a complex model (e.g. DNN with a huge number of model parameters) on a large sensitive genomics datatset distributed across dozens of clients in IID configuration. Additionally, there are a few computing parties in SMPC (practical configuration).
            
            Computational efficiency is an indicator of the extra computational overhead an approach incurs to preserve the privacy. According to Table \ref{tab:comparison} and Figure \ref{fig:comparison}, federated learning is best from this perspective because it follows the paradigm of  ``bringing computation to data``, distributing computational overhead among the clients. HE and SMPC are based on the paradigm of moving data to computation. In HE, encryption of the whole private data in the clients and carrying out computation on encrypted data by the computing party cause a huge amount of overhead. In SMPC, a couple of computing parties process the secret shares from dozens of clients, incurring considerable computational overhead. Among the hybrid approaches, FL+DP has the best computational efficiency given the lower overhead of the two approaches whereas FL+HE has the highest overhead because aggregation process on encrypted parameters is computationally expensive.
            
            Network communication efficiency indicates how efficient an approach utilizes the network bandwidth. The less data traffic is exchanged in the network, the more communication-efficient is the approach. Federated learning is the least efficient approach from the communication aspect since exchanging a large number of model parameter values between the clients and the server generates a huge amount of network traffic. Notice that network bandwidth usage of federated learning is independent of the clients' data because federated learning does not move data to computation but depends on the model complexity (i.e. the number of model parameters). The next approach in this regard is SMPC, where not only each participant sends a large traffic (almost as big as its data) to each computing party but also each computing party exchanges intermediate results (which might be large) with the other computing parties through the network. The network overhead of homomorphic encryption comes from sharing the encrypted data of the clients (as big as the data itself) with the computing party, which is small compared to network traffic generated by federated learning and SMPC. The best approach is differential privacy with no network overhead. Accordingly, FL+DP and FL+SMPC are the best and worst among the hybrid approaches from communication efficiency viewpoint, respectively.
            
            Accuracy of the model in a privacy-preserving approach is a crucial factor in whether to adopt the approach. In the assumed  configuration, SMPC and federated learning are the most accurate approaches incurring little accuracy loss in the final model. Next is homomorphic encryption whose accuracy loss is due to approximating the non-linear operations using addition and multiplication (e.g. least squares approximation \cite{kim2018secure}). The worst approach is differential privacy where the added noise can considerably affect the model accuracy. In the hybrid approaches, FL+SMPC is the best and FL+DP is the worst considering the accuracy of SMPC and differential privacy approaches.
            
            The rest of the comparison measures are privacy-related. The traffic transferred from the clients (participants) to the server (computing parties) is highly sensitive if it carries the private data of the clients. The less sensitive the exchanged traffic is, the more robust the approach is from the privacy perspective. HE and SMPC send the encrypted and anonymous form of the clients' private data to the server, respectively. Federated learning and hybrid approaches share only the model parameters with the server. In HE, if the server has the key to decrypt the traffic from the clients, the whole private data of the clients will be revealed. The same holds if the computing parties in SMPC collude with each other. This might or might not be the case for the other approaches (e.g. federated learning) depending on the exchanged model parameters and whether they can be abused to infer the underlying private data.  
                        
            Privacy of the exchanged traffic indicates how much the traffic is kept private from the server. In HE/SMPC, the data is encrypted/anonymized first and then shared with the server, which is reasonable since it is the clients' private data. In federated learning, the traffic (model parameters) is directly shared with the server assuming that it does not reveal any details regarding individual samples in the data. The aim of the hybrid approaches is to hide the real values of the model parameters from the server to minimize the possibility of inference attacks using the model parameters. FL+HE is the best among the hybrid approaches from this viewpoint.
            
            Privacy guarantee is a metric which quantifies the degree to which the privacy of the clients' data can be preserved. Differential privacy and the corresponding hybrid approach (FL+DP) are the only approaches providing a privacy guarantee, whereas all other approaches can only protect the privacy under a set of certain assumptions. HE assumes that the server does not have the decryption key; The underlying assumption in SMPC is that the computing parties do not collude with each other; federated learning supposes that the model parameters do not give any detail about a sample in the clients' data.
            
        \section*{Discussion and Open Problems}\label{sec:discussion}
            
            From a practical point of view, homomorphic encryption and SMPC that follow the paradigm of ``move data to computation`` do not scale as the number of clients or data size in clients become large. This is because they put the computational burden on a single or a few computing parties. Federated learning, on the other hand, distributes the computation across the clients (aggregation on the server is not computationally heavy) but the communication overhead between the server and clients is the major challenge to scalability of federated learning. The hybrid approaches inherit this issue and it is exacerbated in FL+SMPC. Combining homomorphic encryption with federated learning (FL+HE) adds another obstacle (computational overhead) to scalability of federated learning. There is a growing body of literature on communication-efficient approaches to federated learning that can dramatically improve the scalability of federated learning and make it suitable for large-scale applications including those in biomedicine.
            
            Given that federated learning is the most realistic approach from a scalability viewpoint, it can be used as a standalone approach as long as inferring the clients' data from the model parameters is practically impossible. Otherwise, it should be combined with differential privacy to avoid possible inference attacks and exposure of clients' private data and to provide privacy guarantee. The accuracy of the model will be satisfactory in federated learning but it might be deteriorated in FL+DP. A realistic trade-off needs to be considered depending on the application of interest.
            
            Moreover, differential privacy can have many practical applications in biomedicine as a standalone approach. It works very well for low-sensitivity queries such as counting queries (e.g number of patients with a specific disease) on genomic databases and their generalizations (e.g. histograms) since the presence or absence of an individual changes the query's response by at most one. Moreover, it can be employed to release summary statistics of GWAS such as $\chi^2$ and p-values in a differentially private manner while keeping the accuracy acceptable. A novel  promising research direction is to incorporate differential privacy in deep generative models to generate synthetic genomic and biomedical data.
            
             Future studies can investigate how to reach a compromise between scalability, privacy, and accuracy in real-world settings. The communication overhead of federated learning is still an open problem since although state-of-the-art approaches considerably reduce the network overhead, they adversely affect the accuracy of the model. Hence, novel approaches are required to preserve the accuracy, which is of great importance in biomedicine, while making federated learning communication-efficient.
            
            Adopting federated learning in non-IID settings, where genomic and biomedical datasets across different hospitals/medical centers are heterogeneous, is another important challenge to address. This is because typical aggregation procedures such as simple averaging do not work well for these settings, yielding inaccurate models. Hence, new aggregation procedures are required to tackle non-IID scenarios.
            Moreover, current communication-efficient approaches which were developed for an IID setting might not be applicable to heterogeneous scenarios. Consequently, new techniques are needed to reduce network overhead in these settings, while keeping the model accuracy satisfactory. 
            
            Combining differential privacy  with federated learning to enhance privacy and to provide a privacy guarantee is still a challenging issue in the field. It becomes even more challenging for healthcare applications, where accuracy of the model is of crucial importance. Moreover, the concept of privacy guarantee in differential privacy has been defined for local settings. In distributed scenarios, a dataset might be employed multiple times to train different models with various privacy budgets. Therefore, a new formulation of privacy guarantee should be proposed for distributed settings.
            
        \section*{Conclusions} \label{sec:conclusion}
            
            For AI techniques to succeed, big biomedical data needs to be available and accessible. However, the more AI models are trained on sensitive biological data, the more the awareness about the privacy issues increases, which, in turn, necessitate strategies for shielding the data \cite{berger2019cryptlimit}. Hence, privacy-enhancing techniques are crucial to allow AI to benefit from the sensitive biological data.
            
            Cryptographic techniques, differential privacy and federated learning can be considered as the prime strategies for protecting personal data privacy. These emerging techniques are based on either securing sensitive data, perturbing it or not moving it off site. In particular, cryptographic techniques securely share the data with a single (HE) or multiple computing parties (SMPC), differential privacy adds noise to sensitive data and quantifies privacy loss accordingly, while federated learning enables collaborative learning under orchestration of a centralized server without moving the private data outside local environments. 
            
            All of these techniques have their own strengths and limitations. HE and SMPC are more communication efficient compared to federated learning but they are computationally expensive since they move data to computation and put the computational burden on a server or a few computing parties. Federated learning, on the other hand, distributes computation across the clients but suffers from high network communication overhead. Differential privacy is an  efficient approach from a computational and a communication perspective but it introduces accuracy loss by adding noise to data or model parameters. Hybrid approaches are studied to combine the advantages or to overcome the disadvantages of the individual techniques. 
            We argued that federated learning as a standalone approach or in combination with differential privacy is the most realistic approach to be adopted in biomedicine. We discussed the open problems and challenges in this regard including the balance of communication efficiency and model accuracy in non-IID settings, and need for a new notion of privacy guarantee for distributed biomedical datasets.
            
            Incorporating privacy into the analysis of genomic and biomedical data is still an open challenge, yet preliminary accomplishments are promising to bring practical privacy even closer to real-world settings. Future research should investigate how to achieve a trade-off between scalability, privacy, and accuracy in real biomedical applications.
            
        \section*{Acknowledgements} \label{sec:acc}
        
            This work has received funding from the European Union's Horizon2020 research and innovation program under grant agreement nr. 826078 (FeatureCloud). The work of JB, HS and TK was also supported by Horizon2020 project REPO-TRIAL (nr. 777111). ML, TK, and JB have further been supported by BMBF project Sys\_CARE (01ZX1908A). ML and JB were also supported by BMBF project SYMBOD (01ZX1910D). JB's contribution was also supported by his VILLUM Young Investigator grant (nr. 13154).This paper reflects only the authors' view and the Commission is not responsible for any use that may be made of the information it contains. Figure~\ref{fig:techniques} and Figure~\ref{fig:GM} have been created with BioRender.com. 
            
         \printbibliography
        
        \end{multicols}
\end{document}